\documentclass[sn-nature]{sn-jnl}%

\makeatletter
\let\jnl@style=\rm
\def\ref@jnl#1{{\jnl@style#1}}

\def\aj{\ref@jnl{AJ}}                   %
\def\actaa{\ref@jnl{Acta Astron.}}      %
\def\araa{\ref@jnl{ARA\&A}}             %
\def\apj{\ref@jnl{ApJ}}                 %
\def\apjl{\ref@jnl{ApJ}}                %
\def\apjs{\ref@jnl{ApJS}}               %
\def\ao{\ref@jnl{Appl.~Opt.}}           %
\def\apss{\ref@jnl{Ap\&SS}}             %
\def\aap{\ref@jnl{A\&A}}                %
\def\aapr{\ref@jnl{A\&A~Rev.}}          %
\def\aaps{\ref@jnl{A\&AS}}              %
\def\azh{\ref@jnl{AZh}}                 %
\def\baas{\ref@jnl{BAAS}}               %
\def\bac{\ref@jnl{Bull. astr. Inst. Czechosl.}}
\def\caa{\ref@jnl{Chinese Astron. Astrophys.}}
\def\cjaa{\ref@jnl{Chinese J. Astron. Astrophys.}}
\def\icarus{\ref@jnl{Icarus}}           %
\def\jcap{\ref@jnl{J. Cosmology Astropart. Phys.}}
\def\jrasc{\ref@jnl{JRASC}}             %
\def\memras{\ref@jnl{MmRAS}}            %
\def\mnras{\ref@jnl{MNRAS}}             %
\def\na{\ref@jnl{New A}}                %
\def\nar{\ref@jnl{New A Rev.}}          %
\def\pra{\ref@jnl{Phys.~Rev.~A}}        %
\def\prb{\ref@jnl{Phys.~Rev.~B}}        %
\def\prc{\ref@jnl{Phys.~Rev.~C}}        %
\def\prd{\ref@jnl{Phys.~Rev.~D}}        %
\def\pre{\ref@jnl{Phys.~Rev.~E}}        %
\def\prl{\ref@jnl{Phys.~Rev.~Lett.}}    %
\def\pasa{\ref@jnl{PASA}}               %
\def\pasp{\ref@jnl{PASP}}               %
\def\pasj{\ref@jnl{PASJ}}               %
\def\rmxaa{\ref@jnl{Rev. Mexicana Astron. Astrofis.}}%
\def\qjras{\ref@jnl{QJRAS}}             %
\def\skytel{\ref@jnl{S\&T}}             %
\def\solphys{\ref@jnl{Sol.~Phys.}}      %
\def\sovast{\ref@jnl{Soviet~Ast.}}      %
\def\ssr{\ref@jnl{Space~Sci.~Rev.}}     %
\def\zap{\ref@jnl{ZAp}}                 %
\def\nat{\ref@jnl{Nature}}              %
\def\iaucirc{\ref@jnl{IAU~Circ.}}       %
\def\aplett{\ref@jnl{Astrophys.~Lett.}} %
\def\apspr{\ref@jnl{Astrophys.~Space~Phys.~Res.}}
\def\bain{\ref@jnl{Bull.~Astron.~Inst.~Netherlands}} 
\def\fcp{\ref@jnl{Fund.~Cosmic~Phys.}}  %
\def\gca{\ref@jnl{Geochim.~Cosmochim.~Acta}}   %
\def\grl{\ref@jnl{Geophys.~Res.~Lett.}} %
\def\jcp{\ref@jnl{J.~Chem.~Phys.}}      %
\def\jgr{\ref@jnl{J.~Geophys.~Res.}}    %
\def\jqsrt{\ref@jnl{J.~Quant.~Spec.~Radiat.~Transf.}}
\def\memsai{\ref@jnl{Mem.~Soc.~Astron.~Italiana}}
\def\nphysa{\ref@jnl{Nucl.~Phys.~A}}   %
\def\physrep{\ref@jnl{Phys.~Rep.}}   %
\def\physscr{\ref@jnl{Phys.~Scr}}   %
\def\planss{\ref@jnl{Planet.~Space~Sci.}}   %
\def\procspie{\ref@jnl{Proc.~SPIE}}   %
\makeatother

\usepackage{graphicx}%
\usepackage{multirow}%
\usepackage{amsmath,amssymb,amsfonts}%
\usepackage{amsthm}%
\usepackage{mathrsfs}%
\usepackage[title]{appendix}%
\usepackage{xcolor}%
\usepackage{textcomp}%
\usepackage{manyfoot}%
\usepackage{booktabs}%
\usepackage{algorithm}%
\usepackage{algorithmicx}%
\usepackage{algpseudocode}%
\usepackage{listings}%
\usepackage{mathtools}
\usepackage{subcaption}
\usepackage{gensymb}
\usepackage[export]{adjustbox}

\usepackage[numbers,sort&compress]{natbib}
\usepackage{csquotes}

\begin{document}
\title[Article Title]{\textbf{Black hole jets on the scale of the Cosmic Web}}

\author[1,2]{\fnm{Martijn S.S.L.} \sur{Oei}}
\author[3]{\fnm{Martin J.} \sur{Hardcastle}}
\author[1,4]{\fnm{Roland} \sur{Timmerman}}
\author[5]{\fnm{Aivin R.D.J.G.I.B.} \sur{Gast}}
\author[6]{\fnm{Andrea} \sur{Botteon}}
\author[2]{\fnm{Antonio C.} \sur{Rodriguez}}
\author[7]{\fnm{Daniel} \sur{Stern}}
\author[8,9]{\fnm{Gabriela} \sur{Calistro Rivera}}
\author[1]{\fnm{Reinout J.} \sur{van Weeren}}
\author[1]{\fnm{Huub J.A.} \sur{R\"ottgering}}
\author[1]{\fnm{Huib T.} \sur{Intema}}
\author[6]{\fnm{Francesco} \sur{de Gasperin}}
\author[2]{\fnm{S.G.} \sur{Djorgovski}}

\affil[1]{\orgdiv{Leiden Observatory}, \orgname{Leiden University}, \orgaddress{\street{Niels Bohrweg 2}, \city{Leiden}, \postcode{2333 CA}, \state{Zuid-Holland}, \country{the Netherlands}}}

\affil[2]{\orgdiv{Cahill Center for Astronomy and Astrophysics}, \orgname{California Institute of Technology}, \orgaddress{\street{1216 E California Blvd}, \city{Pasadena}, \postcode{CA 91125}, \state{California}, \country{the United States}}}

\affil[3]{\orgdiv{Centre for Astrophysics Research}, \orgname{University of Hertfordshire}, \orgaddress{\street{College Lane}, \city{Hatfield}, \postcode{AL10 9AB}, \state{Hertfordshire}, \country{the United Kingdom}}}

\affil[4]{\orgname{Centre for Extragalactic Astronomy}, \orgdiv{Department of Physics}, \orgname{Durham University}, \city{Durham}, \postcode{DH1 3LE}, \country{the United Kingdom}}

\affil[5]{\orgdiv{Somerville College}, \orgname{University of Oxford}, \orgaddress{\street{Woodstock Road}, \city{Oxford}, \postcode{OX2 6HD}, \state{Oxfordshire}, \country{the United Kingdom}}}

\affil[6]{\orgname{INAF--IRA}, \orgaddress{\street{Via P. Gobetti 101}, \postcode{40129} \city{Bologna}, \country{Italy}}}

\affil[7]{\orgdiv{Jet Propulsion Laboratory}, \orgname{California Institute of Technology}, \orgaddress{\street{4800 Oak Grove Drive, Mail Stop 264-789}, \city{Pasadena}, \postcode{CA 91109}, \country{the United States}}}

\affil[8]{\orgname{European Southern Observatory}, \orgaddress{\street{Karl-Schwarzschild-Strasse 2}, \postcode{85748}, \city{Garching bei München}, \country{Germany}}}

\affil[9]{\orgname{Institute of Communications and Navigation, German Aerospace Center (DLR)}, \orgaddress{\city{Wessling}, \country{Germany}}}

\abstract{
Jets launched by supermassive black holes transport relativistic leptons, magnetic fields, and atomic nuclei from the centres of galaxies to their outskirts and beyond.
These outflows embody the most energetic pathway by which galaxies respond to their Cosmic Web environment.
Studying black hole feedback is an astrophysical frontier, providing insights on star formation, galaxy cluster stability, and the origin of cosmic rays, magnetism, and heavy elements throughout the Universe.
This feedback's cosmological importance is ultimately bounded by the reach of black hole jets, and could be sweeping if jets travel far at early epochs.
Here we present the joint LOFAR--uGMRT--Keck discovery of a black hole jet pair extending over $7$ megaparsecs --- the largest galaxy-made structure ever found.
The outflow, seen $7.5$ gigayears into the past, spans two-thirds of a typical cosmic void radius, thus penetrating voids at ${\sim}95\%$ probability.
This system demonstrates that jets can avoid destruction by magnetohydrodynamical instabilities over cosmological distances, even at epochs when the Universe was $15$--$7$ times denser than it is today.
Whereas previous record-breaking outflows were powered by radiatively inefficient active galactic nuclei, this outflow is powered by a radiatively efficient active galactic nucleus, a type common at early epochs.
If, as implied, a population of early void-penetrating outflows existed, then black hole jets could have overwritten the fields from primordial magnetogenesis.
This outflow shows that energy transport from supermassive black holes operates on scales of the Cosmic Web and raises the possibility that cosmic rays and magnetism in the intergalactic medium have a non-local, cross-void origin.
}


\maketitle

\section{Main text}\label{sec1}
Nearly every galaxy harbours a spinning supermassive black hole (SMBH) in its centre.
The episodic infall of dust, gas, and stars is believed to activate the Blandford--Znajek mechanism \citep{Blandford11977}, in which electric and magnetic fields convert black hole spin to kinetic energy carried by electrons and positrons.
These leptons form a pair of jets: collimated, relativistic flows along the spin axis that point away from the galactic centre.
Supported by helical magnetic fields \citep[e.g.][]{Pudritz12012}, the most powerful jets avoid disruption by stellar winds \citep[e.g.][]{Perucho12014} and entrain wind-borne atomic nuclei \citep[e.g.][]{Wykes12015}, while blasting off towards intergalactic space.
These jet-driven outflows comprise the majority of bright sources in the known radio sky.

To clarify the impact of black hole energy transport on the intergalactic medium (IGM), recent studies \citep[e.g.][]{Dabhade12020, Oei12022, Oei12023, Mostert12024} searched for Mpc-scale outflows: Nature's largest, and often most powerful, jet systems.
The International LOFAR Telescope \citep[ILT;][]{vanHaarlem12013} has emerged as the prime instrument for their discovery and characterisation.
Our team systematically scanned the ILT's ongoing northern sky survey at wavelength $\lambda = 2.08\ \mathrm{m}$ both with machine learning and by eye --- the latter with significant contributions from citizen scientists \citep{Hardcastle12023}.
This endeavour has increased the number of known Mpc-scale outflows from a few hundred to over eleven thousand \citep{Mostert12024}.

\begin{figure*}[h!]
    \centering
    \begin{subfigure}{.75\textwidth}
    \centering
        \includegraphics[width=\columnwidth]{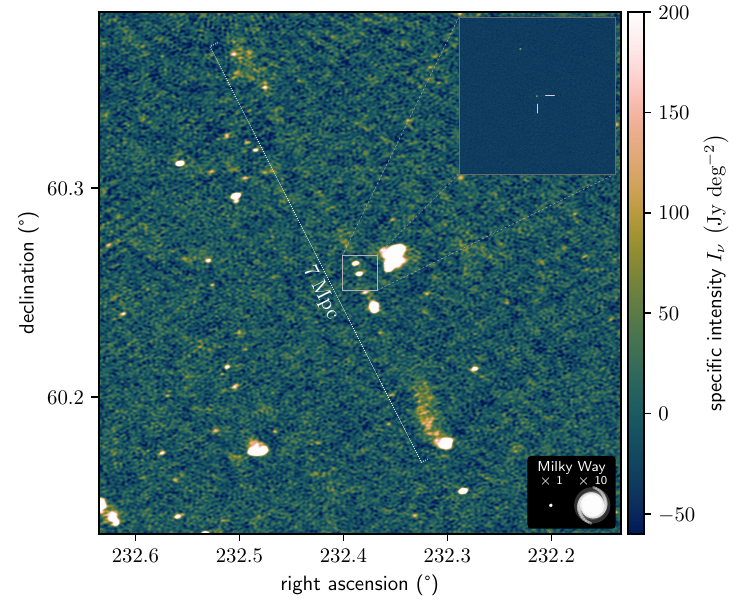}
    \end{subfigure}
    \begin{subfigure}{.75\textwidth}
    \centering
        \includegraphics[width=\columnwidth]{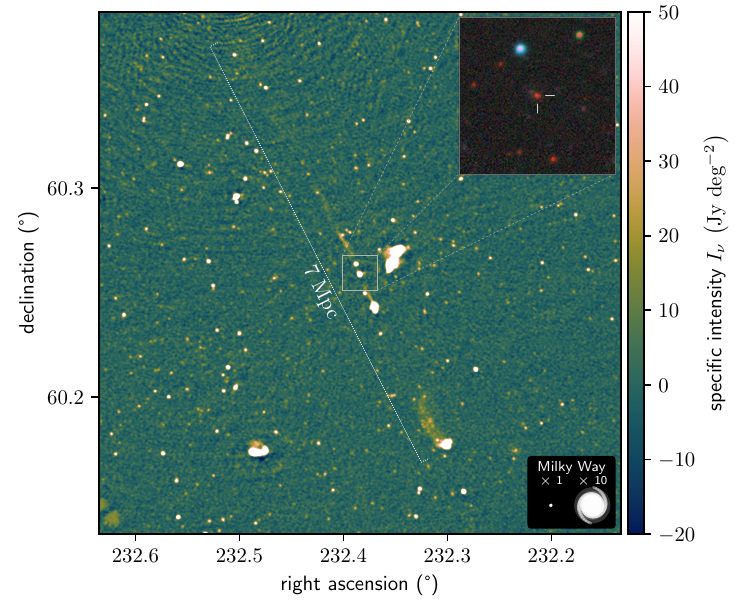}
    \end{subfigure}
    \caption{\textbf{Deep radio images of a 7 Mpc--long, black hole--driven outflow at central wavelengths $\lambda= 2.08\ \mathrm{m}$ (top) and $\lambda = 0.46\ \mathrm{m}$ (bottom).} These images were taken with the ILT and uGMRT, respectively, and have resolutions of $6.2''$ and $4.3''$.
    The top panel's inset shows ILT VLBI imagery at $\lambda = 2.08\ \mathrm{m}$ and a resolution of $0.4''$.
    The bottom panel's inset shows Legacy Survey DR10 optical--infrared imagery.
    The larger images cover $15' \times 15'$ of sky area, whilst the insets cover $1' \times 1'$.
    For scale, we show the stellar Milky Way disk (diameter: 50 kpc) and a ten times inflated version.}
    \label{fig:hypergiantOverview}
\end{figure*}

\begin{figure*}
    \centering
    \begin{subfigure}{\textwidth}
    \centering
    \includegraphics[width=\textwidth]{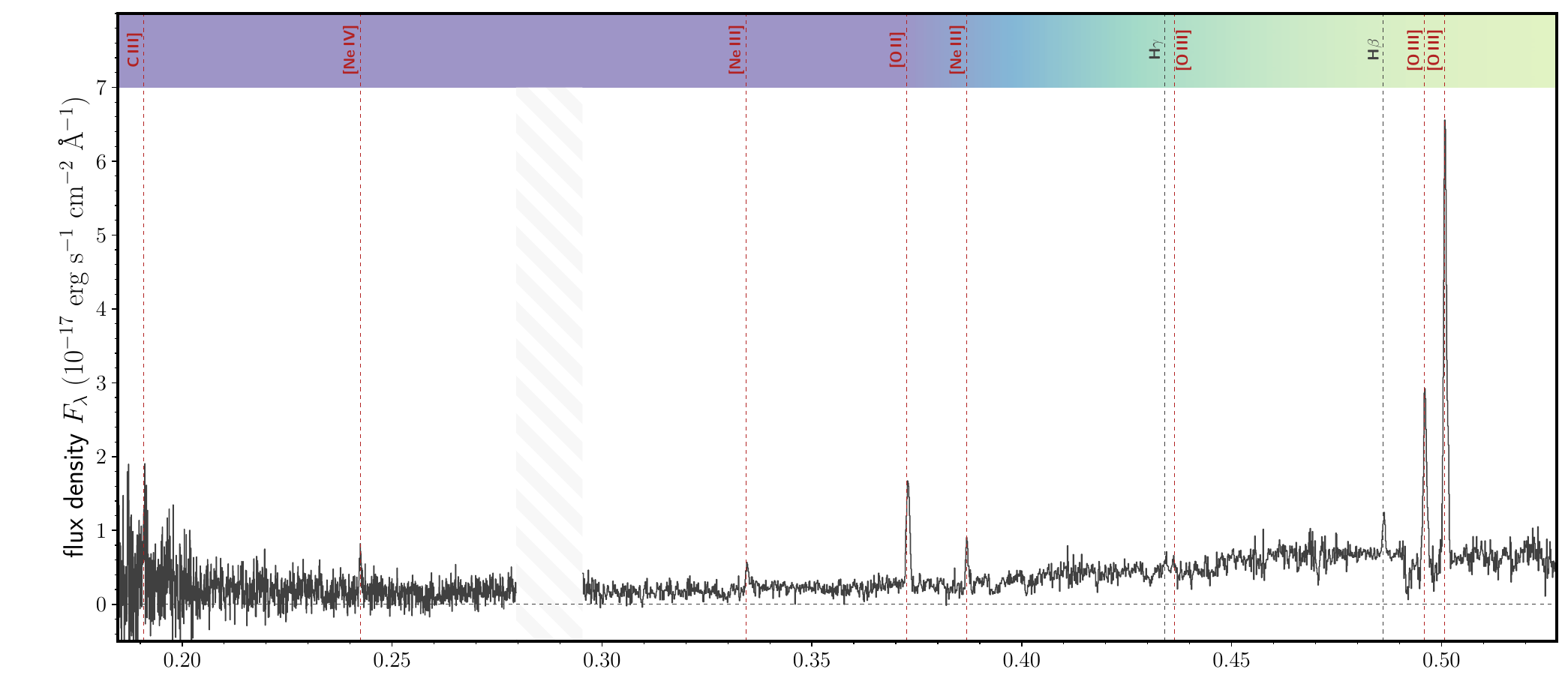}
    \end{subfigure}
    \begin{subfigure}{\textwidth}
    \centering
    \includegraphics[width=\textwidth]{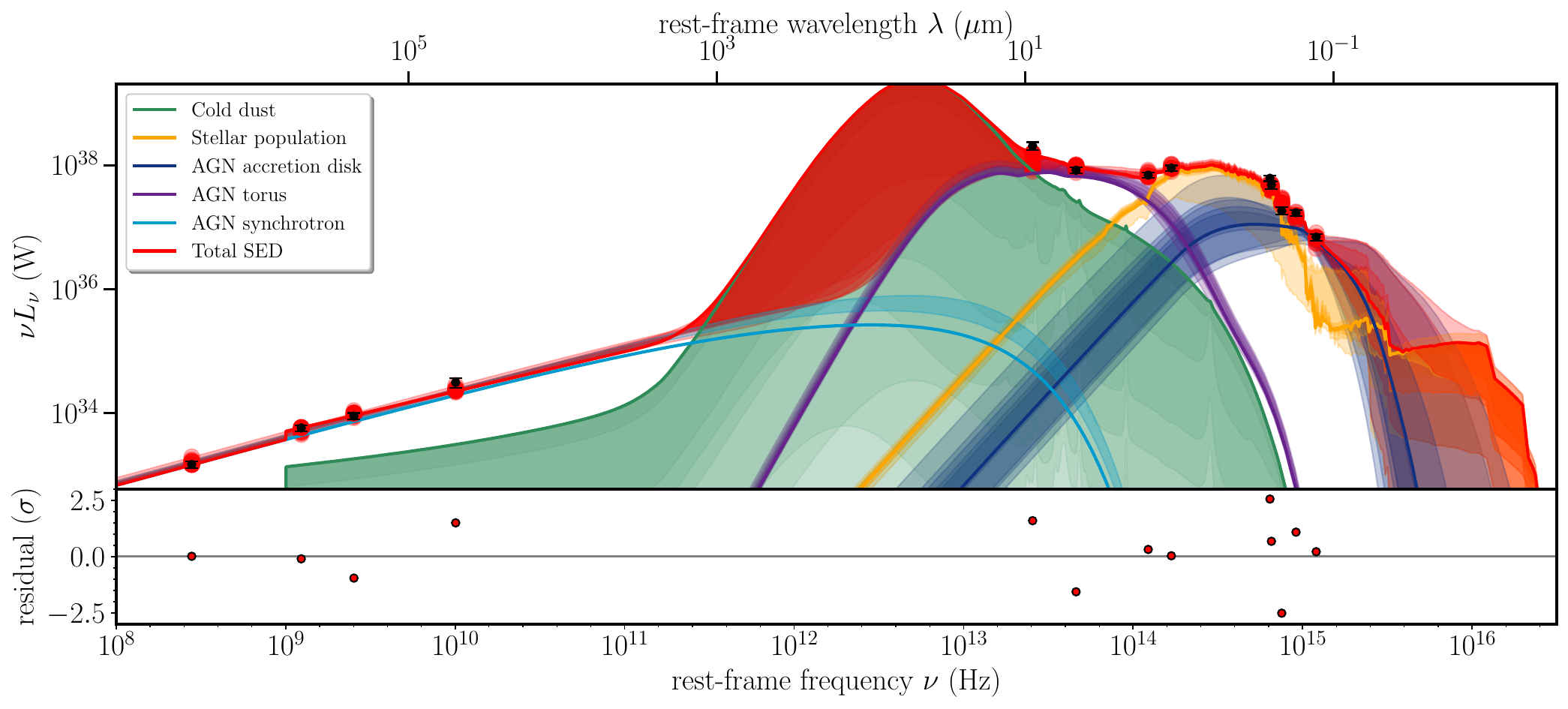}
    \end{subfigure}
    \caption{
    \textbf{
    Both rest-frame ultraviolet--optical spectroscopy (top) and radio--ultraviolet photometry (bottom) demonstrate that the outflow's host galaxy harbours an RE AGN.}
    \textit{Top:} LRIS spectrum exhibiting hydrogen, carbon, oxygen, and neon emission.
    The forbidden lines from multiply ionised oxygen and neon (dark red) could not be generated by even the hottest stars, and instead stem from the narrow-line region of an RE AGN at a redshift $z = 0.896 \pm 0.001$.
    \textit{Bottom:} Bayesian inference of the galaxy's SED (Methods) favours the presence of an AGN accretion disk (dark blue) with an obscuring torus (purple), again indicating radiative efficiency.
    }
    \label{fig:spectrumOpticalHost}
\end{figure*}

Our largest find is the outflow shown in Fig.~\ref{fig:hypergiantOverview}, which we name Porphyrion.
\noindent
The source, of angular length $\phi = 13.4' \pm 0.1'$, is unusually thin.
It consists of a northern lobe, a northern jet, a core, a southern jet with an inner hotspot, and a southern outer hotspot with a backflow.
To investigate from which of two radio-emitting galaxies halfway along the jet axis the outflow originates, we processed ILT very-long-baseline interferometry (VLBI) data of the central $4' \times 4'$.
At a spatial resolution of $3\ \mathrm{kpc}$, the image (Fig.~\ref{fig:hypergiantOverview}'s top panel inset) shows lone, unresolved radio sources in these galaxies, in both cases implying active accretion onto an SMBH.
Because the detection of jets near either black hole (and along the overarching NNE--SSW axis) would clarify Porphyrion's origin, we performed deep follow-up observations with the Upgraded Giant Metrewave Radio Telescope (uGMRT) at $\lambda = 0.46\ \mathrm{m}$.
The resulting image and ancillary optical--infrared data (Fig.~\ref{fig:hypergiantOverview}'s bottom panel) reveal that the outflow protrudes from a massive ($M_\star = 6.7\substack{+1.4\\-1.4} \cdot 10^{11}\ M_\odot$) galaxy.
We observed this galaxy with the Low Resolution Imaging Spectrometer \citep[LRIS;][]{Oke11995, McCarthy11998, Steidel12004, Rockosi12010} on the W.\,M. Keck Observatory's Keck I Telescope, measuring a spectroscopic redshift $z = 0.896 \pm 0.001$ (Fig.~\ref{fig:spectrumOpticalHost}).
\noindent
We witness Porphyrion at $t_\mathrm{BB} = 6.3\ \mathrm{Gyr}$ after the Big Bang.

The outflow's angular length and redshift entail a sky-projected length $l_\mathrm{p} = 6.43 \pm 0.05\ \mathrm{Mpc}$.
This makes Porphyrion the projectively longest known structure generated by an astrophysical body.
The outflow's total length exceeds this projected length, but by how much depends on the unknown inclination of the jets with respect to the sky plane.
Deprojection formulae \citep{Oei12023} predict a total length $l = 6.8\substack{+1.2\\-0.3}\ \mathrm{Mpc}$, with expectation $\mathbb{E}[L\ \vert\ L_\mathrm{p} = l_\mathrm{p}] = 7.28 \pm 0.05\ \mathrm{Mpc}$ (Methods).
We thus estimate Porphyrion to be ${\sim}7\ \mathrm{Mpc}$ long in total.
Spanning ${\sim}66\%$ of the radius of a typical cosmic void at its redshift, the outflow is truly cosmological.
The fact that outflows exceeding 4 Mpc have been known since the 1970s \citep{Willis11974}, whilst those exceeding 5 Mpc remained undiscovered half a century of technological progress later, hitherto suggested a physical limit to outflow growth near 5 Mpc.
Our finding proves this suggestion false.
Surprisingly, SMBH jets can remain collimated over several megaparsecs, despite the growth of (magneto)hydrodynamical (MHD) instabilities --- chiefly Kelvin--Helmholtz instabilities --- predicted theoretically and seen in simulations of shorter jets \citep[e.g.][]{Perucho12019July}.
No MHD simulations of Mpc-scale jets yet exist: the spatio-temporal grids required imply a numerical cost ${\sim}10^2$ times higher than that of state-of-the-art runs.
Outflows like Porphyrion thus offer a window into a jet physics regime that, at present, cannot be explored numerically.

Active galactic nuclei (AGN) with accretion disks extending to the innermost stable circular orbits of their SMBHs efficiently convert the gravitational potential energy of infalling matter into radiation, and are thus called radiatively efficient (RE); all others are called radiatively inefficient (RI) \citep{Heckman12014, Hardcastle12018Comment}.
In RE AGN, the luminous accretion disk photo-ionises a circumnuclear region emitting narrow, and often forbidden, spectral lines.
The Keck-observed prominence of forbidden ultraviolet--optical lines from oxygen and neon (chiefly that of the [O\,III]$\lambda$5007 line, which is $10.3 \pm 0.2$ times brighter than the H$\beta$ line) therefore reveals the presence of an RE AGN \citep{Buttiglione12010}.

By contrast, all previous record-length outflows, such as 3C 236 ($l_\mathrm{p} = 4.6\ \mathrm{Mpc}$; \citep{Willis11974}), J1420--0545 ($l_\mathrm{p} = 4.9\ \mathrm{Mpc}$; \citep{Machalski12008}), and Alcyoneus ($l_\mathrm{p} = 5.0\ \mathrm{Mpc}$; \citep{Oei12022}), are fuelled by RI AGN in recent history ($t_\mathrm{BB} = 10.2$--$12.4\ \mathrm{Gyr}$).
Whereas RI AGN occur primarily in evolved, `red and dead' ellipticals \citep{Heckman12014}, RE AGN feature vigorous gas inflows and are thus generally found in star-forming galaxies.
Indeed, in the first billions of years of cosmic time, RE AGN dominated the radio-bright AGN population \citep{Williams12018}.
The potential of Mpc-scale outflows to spread cosmic rays (CRs), heat, heavy atoms, and magnetic fields through the IGM is particularly high if large specimina could emerge from the type of AGN abundant at early epochs, when the Universe's volume was smaller.
The discovery of a $7\ \mathrm{Mpc}$--long, RE AGN--fuelled outflow before cosmic half-time therefore highlights the hitherto understudied cosmological transport capabilities of Mpc-scale outflows.

\begin{figure*}[t!]
    \centering
    \includegraphics[width=\textwidth]{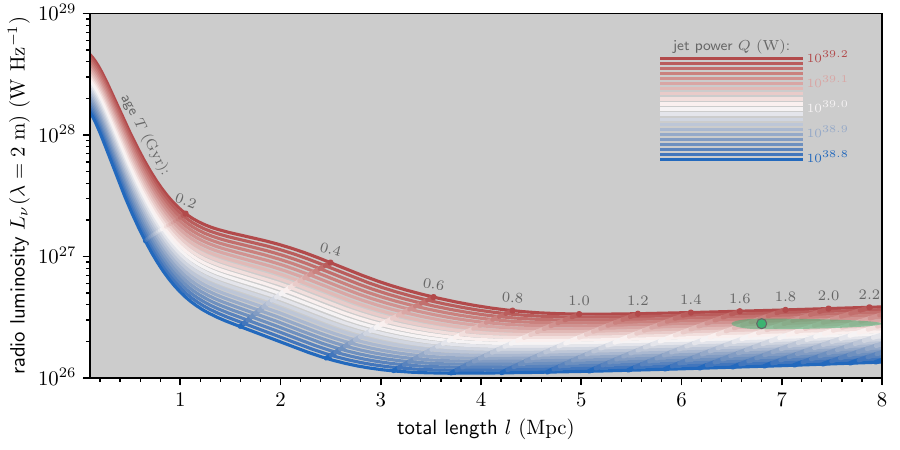}
    \caption{
    \textbf{By superimposing Porphyrion's total length and radio luminosity (green dot) on evolutionary tracks from dynamical modelling (red--white--blue curves), we infer the outflow's two-sided jet power and age.}
    We assume the host galaxy to reside in a galaxy group bordering voids, through which the jets eventually travel.
    }
    \label{fig:inferenceJetPowerAge}
\end{figure*}
The host galaxy likely inhabits a Cosmic Web filament.
Vast voids, which make up the bulk (${\sim}80\%$) of the Universe's volume \citep{Forero-Romero12009}, surround such massive structures in most directions.
Jets as long as Porphyrion's encounter void-like densities and temperatures with high probability (${\sim}95\%$; Methods).
Indeed, the collimated nature of the jets favours scenarios in which they descend into voids, as jets gain resilience against Kelvin--Helmholtz instabilities when the ambient density declines \citep[e.g.][]{Perucho12019July}.
Dynamical modelling suggests a two-sided jet power $Q = 1.3 \pm 0.1 \cdot 10^{39}\ \mathrm{W}$ and an age $T = 1.9\substack{+0.7\\-0.2}\ \mathrm{Gyr}$ (Fig.~\ref{fig:inferenceJetPowerAge}; Methods).
The outflow's average expansion speed $v = 0.012\ c$, comparable to Alcyoneus' \citep{Oei12022}.
In voids and the warm--hot IGM, the speed of sound $c_\mathrm{s} \sim 10^0$--$10^1\ \mathrm{km\ s^{-1}}$: the jets grow hypersonically at Mach numbers $\mathcal{M} \sim 10^2$--$10^3$ and drive strong shocks into voids.
Porphyrion's jets have carried an energy $E = QT = 8\substack{+2\\-1} \cdot 10^{55}\ \mathrm{J}$ into the IGM --- an amount comparable to the energy released during galaxy cluster mergers \citep[e.g.][]{vanWeeren12009}.
This suggests that the outflow is among the most energetic post--Big Bang events to have occurred in its Cosmic Web region.
Even though the SMBH might have gained a significant fraction of its mass while powering the jets ($\Delta M_\bullet > 2\frac{E}{c^2} = 9\substack{+2\\-1} \cdot 10^8\ M_\odot$), it appears to have maintained a constant spin axis throughout gigayears of activity.
Shocks running perpendicular to the jets dissipate enough heat into the filament to increase its temperature by $\Delta T \sim 10^7\ \mathrm{K}$ and its radius by $\Delta r \sim 10^{-1}$--$10^0\ \mathrm{Mpc}$ (Methods).
Outflows like Porphyrion thus locally alter the Cosmic Web's shape.

Figure~\ref{fig:inferenceJetPowerAge} illustrates that the radio luminosity --- and, consequently, the radio surface brightness --- of constant--jet power, Mpc-long outflows decreases over time.
As Fig.~\ref{fig:hypergiantOverview} evinces, Porphyrion borders on the noise of leading current-day telescopes; all outflows further progressed on the same evolutionary track hitherto evade detection.
Similar outflows at higher redshifts or at lower jet powers, and similar but less slenderly shaped outflows, are likewise undetectable.
More generally, statistical modelling \citep{Oei12023} suggests that the detectable population is just the tip of the iceberg: owing to their low radio surface brightnesses, most Mpc-scale outflows are still concealed by noise.
These arguments imply the existence of a hidden population of outflows with sizes comparable to, and possibly larger than, Porphyrion's.

\begin{figure*}[t!]
    \centering
    \includegraphics[width=\textwidth]{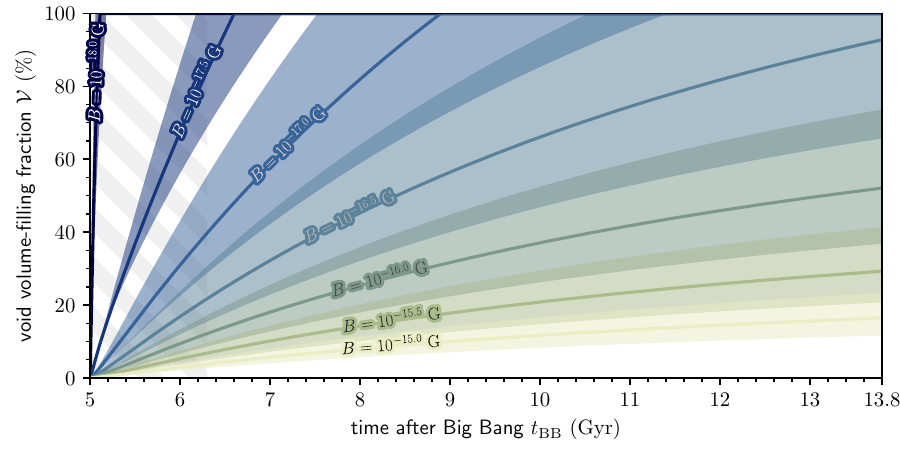}
    \caption{
    \textbf{
    Leptons escaping from the lobes of void-penetrating Mpc-scale outflows diffuse rapidly in weakly magnetised voids. 
    }
    For a single void, and through cosmic time, we show the volumetric fraction filled by electrons and positrons (with $1\ \mathrm{GeV}$ of initial energy) originating from Porphyrion.
    Equally energetic protons diffuse faster, given their minimal inverse Compton losses to the CMB.
    We consider diffusion through turbulent magnetic fields with strengths $B$ and coherence lengths $\lambda_\mathrm{c} \sim 1\ \mathrm{Mpc}$ (spanning a factor two; see translucent bands).
    The hatched strip marks the time prior to Porphyrion's observed state during which its lobes likely (with probability ${>}80\%$) penetrated voids.
    }
    \label{fig:BohmVFF}
\end{figure*}

Mpc-scale outflows long enough to breach filaments, such as Porphyrion, transport large quantities of heavy atoms and CRs into voids \citep{Beck12013}.
In particular, Mpc-scale jets endow $\mathrm{Mpc}^3$-scale volumes in voids with metallicities $Z \sim 10^{-3}$--$10^{-2}\ Z_\odot$ (Methods).
Furthermore, we predict that --- in voids --- the jet- and buoyancy-dominated phases of outflow dynamics are followed by a diffusion phase.
Figure~\ref{fig:BohmVFF} shows the time evolution of the volume-filling fraction of CRs escaping from a void-penetrating lobe.
Many particles undergo this fate: the lobe leaks CR energy at a rate $P \sim 10^{30}\ \mathrm{W} \sim 10^3\ L_\odot$ (Methods), equivalent to a flux of ${\sim}10^{39}$ 1 GeV--particles per second.\footnote{However, in the context of lobe energetics, this loss channel is negligible.
For example, as $P \sim 10^{-9}\ Q$, jet power fluctuations have a far greater effect \citep[e.g.][]{Whitehead12023}.}
The weaker the magnetic fields in voids initially are (see annotations), the greater the mean free path of the diffusing CRs is, and thus the more rapidly they disperse.
If these CRs spread an amount of magnetic energy comparable to their own energy, as suggested by equipartition at source, then a single void-penetrating lobe could fill its void with a magnetic field of strength $B \sim 10^{-16}$--$10^{-15}\ \mathrm{G}$ within a Hubble time (Methods).
Diffusion-driven magnetisation is self-regulating: as the magnetic field strength rises, the mean free path falls, slowing further diffusion.
This mechanism for astrophysical magnetogenesis generates fields consistent with constraints from GeV gamma-ray searches around TeV blazars \citep[e.g.][]{Neronov12010, Chen12015}.

Porphyrion indicates that RE AGN may be at least as effective at generating Mpc-scale outflows as RI AGN are in the Local Universe.
If the comoving number density of actively powered Mpc-scale outflows has remained roughly constant over time at ${\sim}10^1\ (100\ \mathrm{Mpc})^{-3}$ \citep{Oei12023, Mostert12024}, and a comoving volume of $(100\ \mathrm{Mpc})^3$ contains ${\sim}10^1$ voids \citep{Correa12021}, then there would exist ${\sim}1$ actively powered Mpc-scale outflow near every void at every instant.
As Mpc-scale outflows are powered for ${\sim}10^{-2}$--$10^0\ \mathrm{Gyr}$ \citep[e.g.][]{Hardcastle12019, Oei12022}, ${\sim}10^2$ Mpc-scale outflows may have been generated near every void throughout cosmic history.
Only few (${\sim}0.5\%$)\footnote{A single Mpc-scale outflow may penetrate two or more voids.} would need to extend into voids to make CR diffusion from leaky lobes common enough to magnetise the Universe to the observed levels.
Our work suggests that void magnetic fields only trace primordial fields if the latter were strong; otherwise, primordial signals are readily overwritten by void-penetrating Mpc-scale outflows.
Rather than stemming from the Early Universe, magnetism in voids could thus trace the history of black hole energy transport on the scale of the Cosmic Web.

\clearpage

\section{Methods}\label{sec11}
\renewcommand\thefigure{\thesection.\arabic{figure}}
\setcounter{figure}{0}
Throughout this work, we assume a flat, inflationary $\Lambda$CDM cosmological model with parameters from \citet{Planck12020}: $h = 0.6766$, $\Omega_\mathrm{BM,0} = 0.0490$, $\Omega_\mathrm{M,0} = 0.3111$, and $\Omega_{\Lambda,0} = 0.6889$.
We define $\Omega_\mathrm{DM,0} \coloneqq \Omega_\mathrm{M,0}-\Omega_\mathrm{BM,0} = 0.2621$ and $H_0 \coloneqq h \cdot 100\ \mathrm{km\ s^{-1}\ Mpc^{-1}}$.
Furthermore, we define the spectral index $\alpha$ so that it relates to flux density $F_\nu$ at frequency $\nu$ as $F_\nu \propto \nu^{\alpha}$.
Under this convention, synchrotron spectral indices are \emph{positive} (i.e. $\alpha = \frac{5}{2}$) for the lowest frequencies and \emph{negative} for higher frequencies.
As the restoring PSFs may not be perfectly circular, all reported resolutions are effective resolutions.

\paragraph{ILT observations and data reduction}
The International LOFAR Telescope \citep[ILT;][]{vanHaarlem12013} is exquisitely sensitive to the metre-wavelength synchrotron radiation generated by electrons and positrons in the first tens to hundreds of megayears after their acceleration to relativistic energies.
Consequently, the second data release \citep[DR2;][]{Shimwell12022} of the LOFAR Two-metre Sky Survey \citep[LoTSS;][]{Shimwell12017}, the ILT's ongoing northern sky survey in the 120--168~MHz frequency band, has revealed millions of galaxies boasting supermassive black hole (SMBH) jets.

After discovering Porphyrion, the outflow presented in this work, we extracted a total of 16 hours of DDFacet-calibrated visibilities \citep{Tasse12023} from LoTSS pointings P228+60 and P233+60 (Project ID: LT5\_007).
Following \citet{vanWeeren12021}, we subtracted all sources far away from the target, performed phase shifting and averaging, and self-calibrated the resulting data.
This removed residual ionospheric artefacts around ILTJ153004.28+602423.2, the brightest source in the arcminute-scale vicinity of the northern lobe.
We subsequently performed joint multi-scale deconvolution with WSClean \citep{Offringa12014} on the recalibrated target visibilities, yielding the $6.2''$-resolution image of Fig.~\ref{fig:hypergiantOverview}'s top panel.
The noise level is $\sigma = 25\ \mathrm{Jy\ deg^{-2}}$ at its lowest.
The outflow appears thin: its width is nowhere more than a few percent of its length.
We defined Porphyrion's angular length as the largest possible great-circle distance between a point in the southern hotspot and a point in the northern lobe.
The arc connecting these points defines the overarching jet axis, and we measured its position angle to be $27 \pm 1\degree$.

\begin{figure}[t]
\centering
\includegraphics[width=\columnwidth]{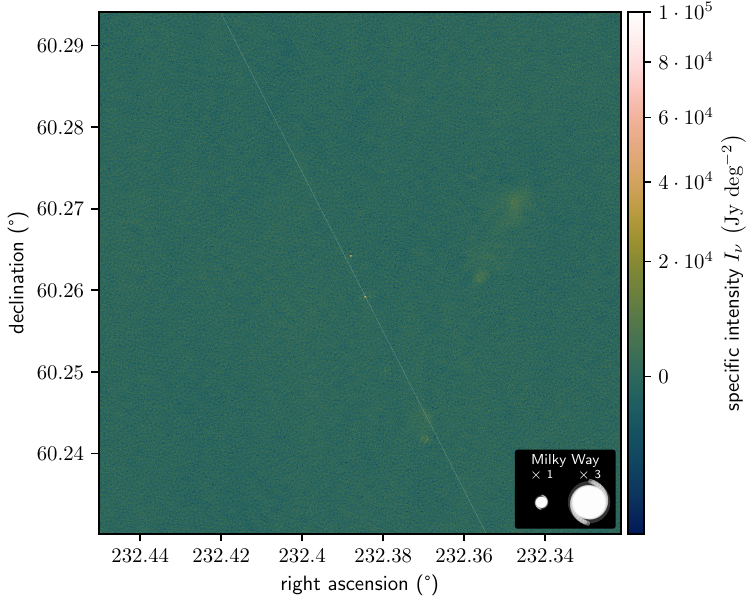}
\caption{\textbf{Our ILT VLBI image of Porphyrion's central $3.84' \times 3.84'$ at $\lambda = 2.08\ \mathrm{m}$ and $0.4''$ resolution covers a third of the total jet system and reveals two radio-luminous AGN.}
We show the overarching jet axis (translucent white), determined from the northern lobe and southern hotspot (not shown), to scale for a jet radius of 1 kpc.
The jet axis appears to pass through J152932.16+601534.4.
}
\label{fig:hypergiantILTVLBI}
\end{figure}

To obtain a higher resolution image of Porphyrion, we reprocessed the P233+60 data, including LOFAR's international stations, from scratch using the LOFAR-VLBI pipeline \citep{Morabito12022}.
This pipeline builds upon the calibration pipeline for the Dutch part of the array to calibrate the international stations.
We derived the dispersive phase corrections and gain corrections for the international stations by calibrating against a bright and compact radio source near the target.
In this case, we used the aforementioned ILTJ153004.28+602423.2, a known source from the Long-Baseline Calibrator Survey \citep[LBCS;][]{Jackson12016, Jackson12022}.
To reduce interference from unrelated radio sources in Porphyrion's angular vicinity, we phased up LOFAR's core stations to narrow down the field of view and only considered data from long baselines to calculate the calibration solutions.
With the calibration solutions applied in the direction of the target, we again performed deconvolution with WSClean to obtain a $0.4''$-resolution image, which we show partially in Fig.~\ref{fig:hypergiantOverview}'s top panel inset and fully in Fig.~\ref{fig:hypergiantILTVLBI}.
The noise level is $\sigma = 2.7 \cdot 10^3\ \mathrm{Jy\ deg^{-2}}$ at its lowest.
This image, which covers the central one-third of the total jet system, reveals synchrotron emission at $42\sigma$ significance from active galactic nuclei (AGN) in only two galaxies, $19''$ apart.
Both lie along the outflow's jet axis nearly halfway between its endpoints.
We considered these galaxies, J152933.03+601552.5 and J152932.16+601534.4, to be Porphyrion's host candidates.
In contrast to other radio-emitting structures along Porphyrion's axis, such as the southern complex interpreted as an inner hotspot, these candidates have optical counterparts in Legacy Surveys DR10 imagery (see Fig.~\ref{fig:hypergiantOverview}'s bottom panel inset).

\paragraph{uGMRT observations and data reduction}
On 13 May 2023, we observed the outflow with the uGMRT in Band 4 (550--750 MHz) for a total of 10 hours.
On 23 September 2023, we extended these observations with another 5 hours.
These observations are part of GMRT Observing Cycle 44 and have project code 44\_101.
We requested to record both narrow-band (GSB) and wide-band (GWB) data.
Adverse ionospheric conditions during the September run prohibited us from improving upon the images produced with the May run data only.
In what follows, we therefore exclusively discuss May run data reduction and results.
We performed calibration with Source Peeling and Atmospheric Modeling \citep[SPAM;][]{Intema12014}, starting out with the GSB data.
After direction-dependent calibration, we used Python Blob Detection and Source Finder \citep[PyBDSF;][]{Mohan12015} to derive a sky model from the final GSB image, which subsequently served to initialise the direction-dependent calibration of the GWB data.
As SPAM was designed with narrow-band data in mind, following standard practice, we first split the GWB data along the frequency axis, yielding four subbands of 50 MHz width each.
We then calibrated each subband independently.
A joint image of four calibrated subbands revealed residual ionospheric artefacts from ILTJ153004.28+602423.2, the same bright source in the vicinity of the northern lobe mentioned earlier.
To mitigate these artefacts, we subtracted (on a subband basis) all sources outside of a spherical cap with a $9'$ radius centred around J2000 right ascension $\varphi = \text{15h29m32.0s}$ and declination $\theta = \text{60d15m33.0s}$.
We then jointly reimaged the four source-subtracted subbands with WSClean, using Briggs weighting 0.
This resulted in the $4.3''$-resolution image of Fig.~\ref{fig:hypergiantOverview}'s bottom panel.
The noise level is $\sigma = 3\ \mathrm{Jy\ deg^{-2}}$ at its lowest.

\begin{figure}
    \centering
    \includegraphics[width=\columnwidth]{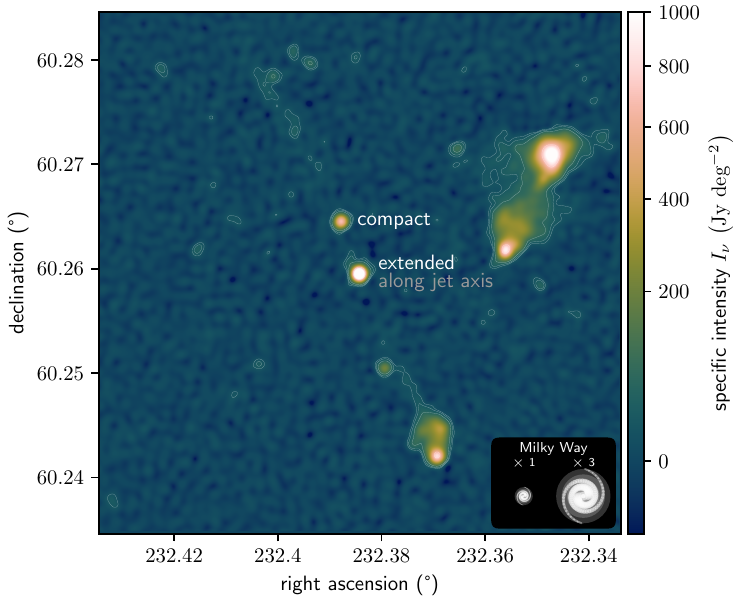}
    \caption{
    \textbf{
    In our imagery, only the southern host galaxy candidate features a radio extension along Porphyrion's overarching jet axis.
    }
    For the central $3' \times 3'$ sky area, we show a uGMRT image at $\lambda = 0.46\ \mathrm{m}$ and $3.6''$ resolution.
    We detect the southern galaxy's radio extension, directed towards the north-northeast, at $5\sigma$ significance.
    The contours denote $3\sigma$, $5\sigma$, $10\sigma$, and $100\sigma$.
    }
    \label{fig:hypergiantCentreStokesI}
\end{figure}

In the Legacy Survey DR10 optical imagery shown in Fig.~\ref{fig:hypergiantOverview}'s bottom panel inset, we identified two faint galaxies in the arcsecond-scale vicinity of the southern host galaxy candidate.
Of these, the galaxy at $(\varphi,\theta) = (232.37969\degree, 60.26029\degree)$ emits low-frequency radio emission at $6\sigma$ significance.
At the $4.3''$ resolution of our fiducial uGMRT image, this radio emission is only narrowly separable from the host galaxy candidate's, thus interfering with establishing the radio morphology of the candidate.
Trading depth for resolution, we reimaged the uGMRT data with WSClean using Briggs weighting $-0.5$, yielding a $3.6''$ resolution.
Subsequently, to isolate the radio morphology of J152932.16+601534.4, we fit a circular Gaussian fixed at the sky coordinates of its radio-emitting neighbour.
Naturally, we set this Gaussian's full width at half maximum to $3.6''$.
Upon subtracting the Gaussian, we obtained our final image; Fig.~\ref{fig:hypergiantCentreStokesI} shows its central region, where the noise level is $\sigma = 6\ \mathrm{Jy\ deg^{-2}}$ at its lowest.
Only the southern (and most radio-luminous) host galaxy candidate features an extension along the overarching jet axis seen in Fig.~\ref{fig:hypergiantOverview}.
In our data, this extension --- indicative of a pair of relativistically beamed jets --- occurs at $5\sigma$ significance.
We conclude that J152932.16+601534.4 is Porphyrion's host galaxy.


\paragraph{Keck I observations and data reduction}
\begin{figure*}[ht]
    \centering
    \includegraphics[width=\textwidth]{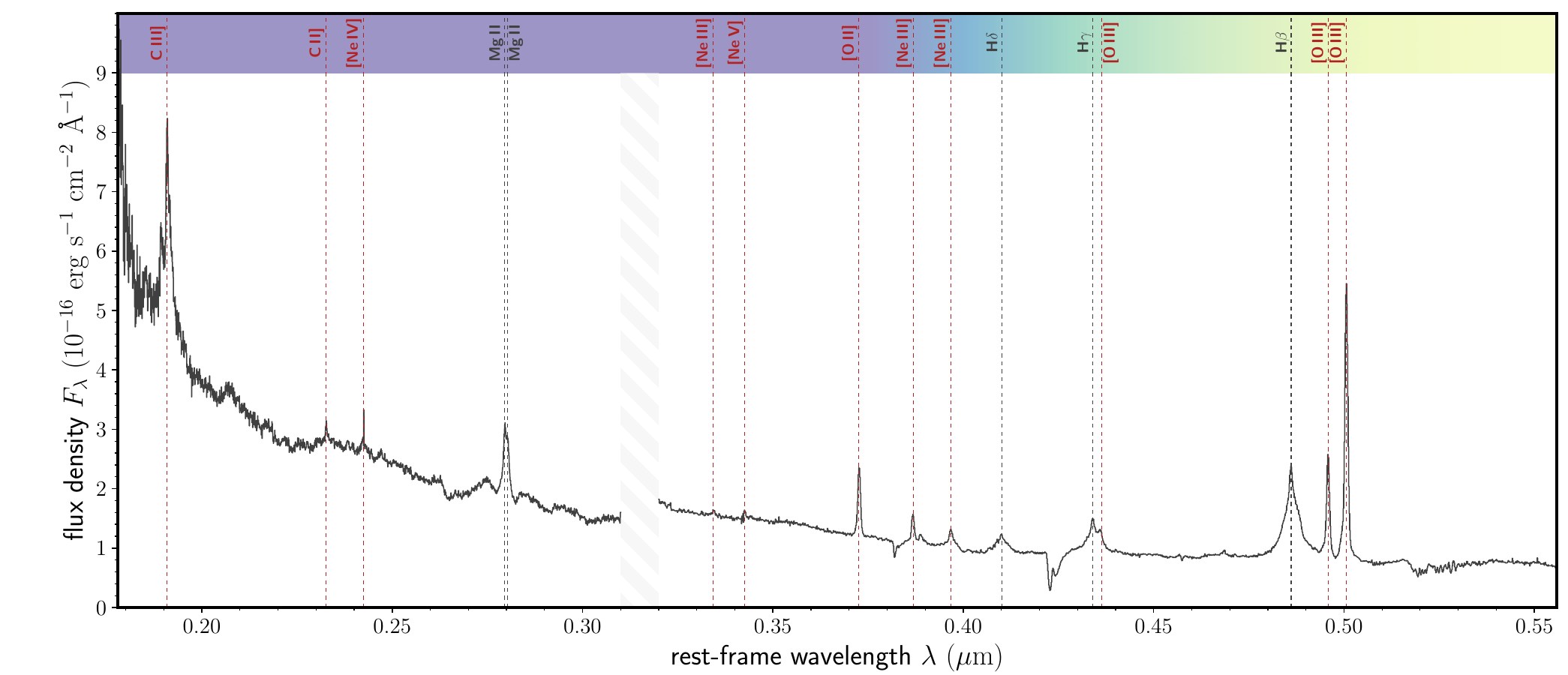}
    \caption{
    \textbf{Ultraviolet--optical rest-frame spectrum of J152933.03+601552.5, the quasar-hosting galaxy $19''$ north-northeast of J152932.16+601534.4, Porphyrion's host galaxy.}
    We identify redshifted hydrogen, carbon, oxygen, neon, and magnesium lines, jointly implying $z_\mathrm{s} = 0.799 \pm 0.001$.
    Forbidden lines from the quasar's narrow-line region are shown in red.
    The spectrum has been measured with the LRIS on the W.\,M. Keck Observatory's Keck I Telescope.
    }
    \label{fig:spectrumOpticalQuasar}
\end{figure*}
The literature offers only photometric redshift estimates of the host galaxy.
The SDSS DR12 \citep{Alam12015} reports $z_\mathrm{p} = 0.68 \pm 0.06$, the Legacy Surveys DR9 \citep{Dey12019} reports $z_\mathrm{p} = 0.93 \pm 0.08$, and \citet{Duncan12022} reports $z_\mathrm{p} = 0.92 \pm 0.08$.
For radio-emitting galaxies like J152932.16+601534.4, we consider the latter estimate to be most reliable.

To establish the redshift of Porphyrion's host galaxy with certainty, we measured its (rest-frame) ultraviolet--optical spectrum with the Low Resolution Imaging Spectrometer \citep[LRIS;][]{Oke11995, McCarthy11998, Steidel12004, Rockosi12010} on the W.\,M. Keck Observatory's Keck I Telescope.
Adequate slit placement requires accurate knowledge of the galaxy's coordinates.
From the Legacy Surveys DR10 best-fit model, we found that J152932.16+601534.4's centre lies at $(\varphi, \theta) = (232.38410 \degree, 60.25960 \degree)$.
The galaxy's half-light radius is $10.1 \pm 0.3\ \mathrm{kpc}$.
On 23 June 2023, we observed the galaxy for a total of 900 seconds.
We used the 600/4000 grism on LRIS' blue side, with $1 \times 2$ binning (spatial and spectral, respectively), and the 400/8500 grating on the red side, again with $1 \times 2$ binning.
During the observations, the seeing was approximately $0.8''$; as we used a $1.5''$ slit, minimal slit losses occurred.
Using a slit position angle of $-70\degree$, we could simultaneously obtain a spectrum for J152933.03+601552.5, the quasar-hosting galaxy which we initially considered (and then discarded) as a host candidate.
We reduced the data with PypeIt \citep{Prochaska12020}, a Python-based pipeline with features tailored to reducing LRIS long-slit spectroscopy.
We flat-fielded and sky-subtracted the data using standard techniques.
We used internal arc lamps for wavelength calibration and a standard star for overall flux calibration.

The final LRIS-derived spectra of J152932.16+601534.4 and J152933.03+601552.5 are shown in Figs.~\ref{fig:spectrumOpticalHost} and \ref{fig:spectrumOpticalQuasar}, respectively.
The corresponding spectroscopic redshifts are $z_\mathrm{s} = 0.896 \pm 0.001$ and $z_\mathrm{s} = 0.799 \pm 0.001$.
The uncertainties reflect LRIS' limited spectral resolution as well as systematic errors in wavelength calibration.
The latter spectroscopic redshift can be compared to the value derived for J152933.03+601552.5 by the SDSS BOSS \citep{Dawson12013} on 5 July 2013.
Visual inspection of the SDSS BOSS spectrum and its best fit indicates a robust spectroscopic redshift $z_\mathrm{s} = 0.79836 \pm 5 \cdot 10^{-5}$.
The two measurements are in agreement.

\paragraph{Spectral energy distribution}
To further assess the accretion mode of Porphyrion's AGN, and to estimate its host's stellar mass and possibly star formation rate (SFR), we performed spectral energy distribution (SED) inference.
Through VizieR, we collected catalogued total (rather than fixed-aperture) flux densities, relative flux densities, and magnitudes from rest-frame ultraviolet to radio wavelengths.
Just ${\sim}3''$ northeast from Porphyrion's host galaxy lies another source, which could be either a Milky Way star or a galaxy.
Mindful of the possibility of spuriously high flux density measurements as a result of target--neighbour blending, we assessed all images underlying the catalogued estimates by eye.
The neighbouring source only appears to be a point of attention for flux density measurements at small wavelengths, such as in the Legacy \textit{g}- and \textit{r}-band, where it has flux densities ${\sim}100\%$ and ${\sim}60\%$ those of the target, respectively.
At the Legacy \textit{z}-band's larger wavelengths, the neighbour's flux density is small (${\sim}20\%$) relative to the target's.
The error induced by blending, which will add only a fraction of the neighbour's flux density, should thus be negligible.
Accordingly, the Pan-STARRS and WISE measurements at even larger wavelengths are not compromised by this neighbour.
We converted the Legacy relative flux densities to flux densities by multiplying with the reference flux density $F_\nu = 3631\ \mathrm{Jy}$.
We converted the Pan-STARRS AB magnitudes to flux densities using the standard relation (e.g. Eq. 1 of \citet{Chambers12016}).
We converted the WISE relative flux densities to flux densities by multiplying with the reference flux densities of \citet{Jarrett12011}'s Table 1.
Table~\ref{tab:fluxDensities} provides all retained flux densities $F_\nu$ and the central wavelengths $\lambda$ they correspond to.
\begin{table}[b]
\caption{
Porphyrion's host galaxy flux densities $F_\nu$ throughout the electromagnetic spectrum.
Entries are sorted by the central wavelengths $\lambda$ of the observing bands.$^1$
}%
\label{tab:fluxDensities}
\begin{tabular*}{\columnwidth}{@{\extracolsep\fill}lll}
\toprule%
Band & $\lambda\ (\mu\mathrm{m})$ & $F_\nu\ (\mathrm{Jy})$\\
\midrule
Legacy \textit{g}  & $4.8 \cdot 10^{-1}$ & $2.6 \pm 0.2 \cdot 10^{-6}$\\
Legacy \textit{r}  & $6.3 \cdot 10^{-1}$ & $8.4 \pm 0.4 \cdot 10^{-6}$\\
Legacy \textit{z}  & $9.1 \cdot 10^{-1}$ & $4.31 \pm 0.08 \cdot 10^{-5}$\\
Pan-STARRS \textit{i} & $7.5 \cdot 10^{-1}$ & $1.1 \pm 0.1 \cdot 10^{-5}$\\
Pan-STARRS \textit{y} & $9.6 \cdot 10^{-1}$ & $3.3 \pm 0.3 \cdot 10^{-5}$\\
WISE W1 & $3.4 \cdot 10^{0}$ & $2.41 \pm 0.02 \cdot 10^{-4}$\\
WISE W2 & $4.6 \cdot 10^{0}$ & $2.53 \pm 0.05 \cdot 10^{-4}$\\
WISE W3 & $1.2 \cdot 10^{1}$ & $8.1 \pm 0.5 \cdot 10^{-4}$\\
WISE W4 & $2.2 \cdot 10^{1}$ & $3.6 \pm 0.4 \cdot 10^{-3}$\\
VLASS & $1.0 \cdot 10^{5}$ & $1.4 \pm 0.2 \cdot 10^{-3}$\\
FIRST & $2.1 \cdot 10^{5}$ & $1.6 \pm 0.1 \cdot 10^{-3}$\\
uGMRT Band 4 & $4.6 \cdot 10^{5}$ & $2.1 \pm 0.1 \cdot 10^{-3}$\\
LoTSS & $2.1 \cdot 10^{6}$ & $2.4 \pm 0.2 \cdot 10^{-3}$\\
\botrule
\end{tabular*}
\footnotetext[1]{
\footnotesize
When multiple flux densities or magnitudes from the same band were available in literature catalogues, we picked the highest signal-to-noise ratio measurement.
Legacy data come from \citet{Dey12019}, Pan-STARRS data from \citet{Chambers12016}, WISE data from \citet{Lang12016}, VLASS data from \citet{Gordon12021}, FIRST data from \citet{Helfand12015}, uGMRT data from the present work, and LoTSS data from \citet{Shimwell12022}.}
\end{table}

Next, using AGNfitter \citep[][Mart\'inez-Ram\'irez et al. in prep.]{CalistroRivera12016}, we determined the SED posterior shown in the bottom panel of Fig.~\ref{fig:spectrumOpticalHost}.
The posterior indicates the presence of a luminous SMBH accretion disk with an obscuring torus, confirming the radiatively efficient nature of Porphyrion's AGN.
Our model requires the disk and torus to explain the observed infrared (WISE) and near-ultraviolet (Legacy) flux levels, which exceed those possible with cold dust and stars alone.

The SED posterior further implies that the stellar mass of Porphyrion's host is $M_\star = 6.7 \pm 1.4 \cdot 10^{11}\ M_\odot$.
To gauge the sensitivity of stellar mass estimates for this galaxy to methodological variation, we compare our result to the corresponding stellar mass estimate in the LoTSS DR2 value-added catalogue \citep{Hardcastle12023}.
This catalogue's authors derive a stellar mass $M_\star = 5.5\substack{+0.7\\-0.6} \cdot 10^{11}\ M_\odot$ from SED fits to Legacy $g$, $r$, $z$ and WISE W1 and W2 flux densities.\footnote{This stellar mass estimate is not based on the spectroscopic redshift we have obtained through LRIS, but utilises a photometry-based redshift posterior with mean and standard deviation $z_\mathrm{p} = 0.92 \pm 0.08$ \citep{Duncan12022}.}
The two stellar mass measurements are in agreement.
Due to the lack of rest-frame far-infrared photometry, the SFR of Porphyrion's host is virtually unconstrained by the SED posterior.

\begin{figure}
    \centering
    \begin{subfigure}{.7\columnwidth}
        \centering
        \includegraphics[width=\columnwidth]{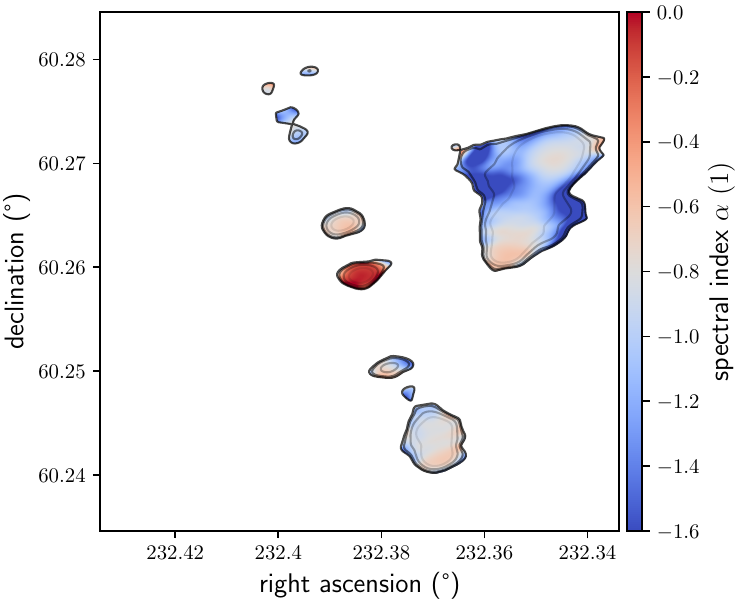}
    \end{subfigure}
    \begin{subfigure}{.7\columnwidth}
        \centering
        \includegraphics[width=\columnwidth]{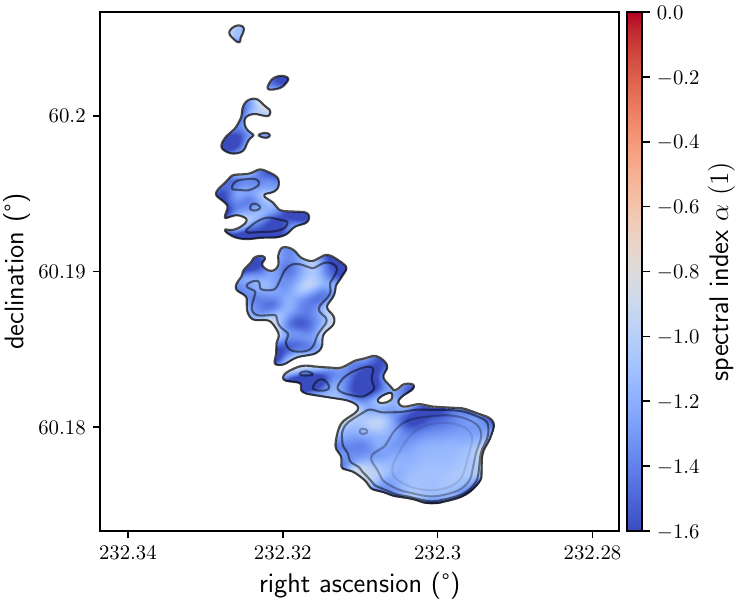}
    \end{subfigure}
    \caption{
    \textbf{Metre-wavelength spectral indices around Porphyrion's centre and southern tip.}
    The top panel, which covers $3' \times 3'$, reveals synchrotron self-absorption at metre wavelengths in the host galaxy, consistent with the fuelling of powerful jets.
    The bottom panel, which covers $2' \times 2'$, reveals a hotspot with backflow.
    We show the mean spectral index $\alpha$ between 0.46--2.08 m, at a resolution of $6.2''$.
    From light to dark, the contours denote thermal noise--induced spectral index uncertainties of 0.05, 0.1, 0.2, and 0.3.
    }
    \label{fig:hypergiantSpectralIndexMaps}
\end{figure}
\paragraph{Radio luminosities and spectral indices}
To determine metre-wavelength radio luminosities and a metre-wavelength spectral index for Porphyrion, we first measured its flux densities in the $6.2''$ ILT and $4.3''$ uGMRT images.
We assumed flux scale uncertainties of $10\%$ and $5\%$, respectively.
Summing over all structural components, the outflow's total flux density at $\lambda = 2.08\ \mathrm{m}$ is $F_\nu = 63 \pm 6\ \mathrm{mJy}$; its total radio luminosity at rest-frame wavelength $\lambda_\mathrm{r} = 1.10\ \mathrm{m}$ therefore is $L_\nu = 1.4 \pm 0.1 \cdot 10^{26}\ \mathrm{W\ Hz^{-1}}$.
The outflow's total flux density at $\lambda = 0.46\ \mathrm{m}$ is $F_\nu = 12.0 \pm 0.6\ \mathrm{mJy}$; its total radio luminosity at rest-frame wavelength $\lambda_\mathrm{r} = 0.24\ \mathrm{m}$ therefore is $L_\nu = 2.7 \pm 0.1 \cdot 10^{25}\ \mathrm{W\ Hz^{-1}}$.
These data imply a metre-wavelength spectral index $\alpha = -1.09 \pm 0.08$.
Through spectral index--based interpolation, we estimated the total radio luminosity at rest-frame wavelength $\lambda_\mathrm{r} = 2\ \mathrm{m}$ to be $L_\nu = 2.8 \pm 0.3 \cdot 10^{26}\ \mathrm{W\ Hz^{-1}}$.
This latter total radio luminosity is an important input for our dynamical modelling.

We calculated directionally resolved metre-wavelength spectral indices by combining the ILT and uGMRT images.
Before doing so, we convolved the latter image to the former's resolution.
In Fig.~\ref{fig:hypergiantSpectralIndexMaps}, we show two regions of interest from the resulting spectral index map, which consequently has a resolution of $6.2''$.
To highlight the directions in which our spectral index measurements are informative, we blanked all directions in which the thermal noise--induced spectral index uncertainty exceeds 0.3.
The top panel of Fig.~\ref{fig:hypergiantSpectralIndexMaps} shows that J152932.16+601534.4, Porphyrion's host galaxy, has a significantly higher spectral index than J152933.03+601552.5, the aforementioned quasar-hosting galaxy.
The former spectral index is consistent with zero, indicating that the onset of synchrotron self-absorption (SSA) in Porphyrion's host galaxy occurs at metre wavelengths.
By contrast, the onset of SSA in the quasar-hosting galaxy must occur at longer wavelengths, suggesting a lower lepton energy density and weaker magnetic fields in its synchrotron-radiating region.
The bottom panel of Fig.~\ref{fig:hypergiantSpectralIndexMaps} shows that Porphyrion's southern tip features much lower spectral indices, with a gradient along the jet axis.
This gradient is consistent with a scenario of a hotspot with backflow in which spectral ageing occurs.
Whereas $\alpha = -1.0 \pm 0.2$ at the hotspot's southwestern side, the radio spectra gradually steepen to $\alpha = -1.6 \pm 0.2$ at the hotspot's northeastern side.
No spectral trend appears present further downstream.

\begin{figure}
    \centering
    \includegraphics[width=\columnwidth]{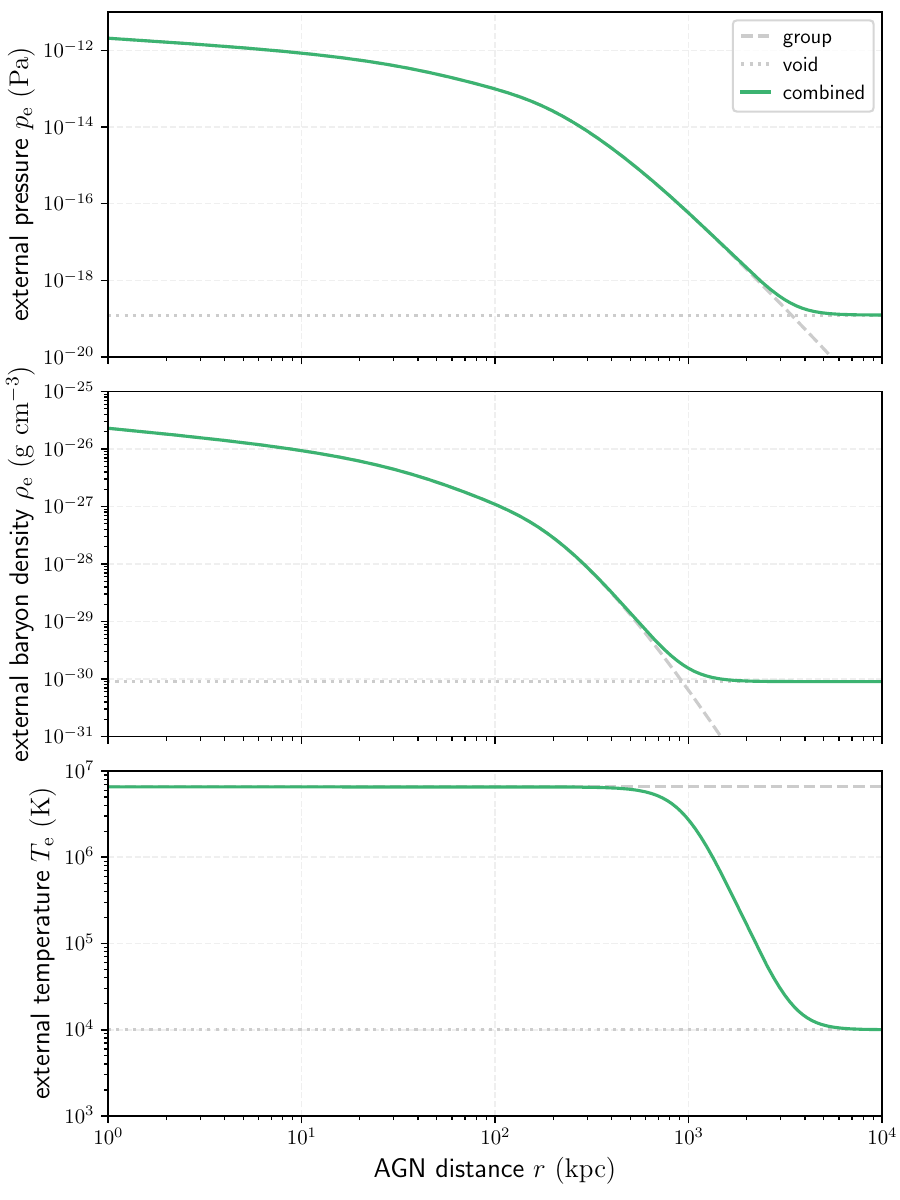}
    \caption{
    \textbf{Pressure, baryon density, and temperature external to the outflow, as a function of the proper distance from Porphyrion's AGN, in our dynamical modelling.}
    The profiles consist of contributions from the outflow's presumed galaxy group and the adjacent voids.
    }
    \label{fig:dynamicalModel}
\end{figure}

\paragraph{Dynamical modelling: jet power and age}
We derived Porphyrion's jet power and age from its length, radio luminosity, cosmological redshift, and likely environment by fitting evolutionary tracks.
We generated these evolutionary tracks with the simulation-based analytic outflow model of \citet{Hardcastle12018}.
This model requires assumptions on the large-scale environment in which the dynamics take place.
The Legacy Survey DR10 optical imagery in the inset of Fig.~\ref{fig:hypergiantOverview}'s bottom panel suggests that the host galaxy does not reside in a galaxy cluster, which would be visible at this redshift as a galaxy overdensity on the sky.
Concordantly, studies have found that jet-fuelling RE AGN avoid rich environments \citep{Ineson12013, Ineson12015}.
Nevertheless, the straightness of the outflow implies a low peculiar speed ($v_\mathrm{p} \lesssim 10^2\ \mathrm{km\ s^{-1}}$), and consequently that the host galaxy is at the bottom of a local gravitational potential well.
We thus instead suppose that the host galaxy resides in the centre of a galaxy group of mass $M_{500} = 10^{13}\ M_\odot$ (which comprises contributions from both dark and baryonic matter) \citep{Pasini12021, Oei12024}.
We assigned the group a universal pressure profile \citep[UPP;][]{Arnaud12010} $p_\mathrm{g}(r)$,\footnote{\citet{Sun12011} have shown that the UPP applies to galaxy groups, even though the profile has originally been proposed to fit data on galaxy \emph{clusters} (which have much higher masses: $10^{14}\ M_\odot < M_{500} < 10^{15}\ M_\odot$).} which can be parametrised just by $M_{500}$.
To obtain the group's baryon density profile from its pressure profile, we invoked the ideal gas law: $\rho_\mathrm{g}(r) = \frac{p_\mathrm{g}(r) \langle m \rangle}{k_\mathrm{B}T_\mathrm{g}}$, where $\langle m \rangle$ is the average plasma particle mass and $T_\mathrm{g}$ the group temperature.
We assumed a pure $^1\mathrm{H}$--$^4\mathrm{He}$ plasma with a $^4\mathrm{He}$ mass fraction $Y = 25\%$ \citep[e.g.][]{Cooke12018}, so that $\langle m \rangle \approx \frac{4}{8-5Y} m_\mathrm{p} = 0.6\ m_\mathrm{p}$, where $m_\mathrm{p}$ is the proton mass.
We estimated $T_\mathrm{g}$, which we assumed constant in space and time, using the mass--temperature relation specified by Eq.~9 and Tables~3 and 4 of \citet{Lovisari12015}:
\begin{align}
\frac{k_\mathrm{B}T_\mathrm{g}}{2\ \mathrm{keV}} = 0.77 \cdot \left(\frac{M_{500}}{5 \cdot 10^{13}\ h_{70}^{-1}\ M_\odot}\right)^{0.61}.
\end{align}
The aforementioned mass implies $T_\mathrm{g} = 7 \cdot 10^6\ \mathrm{K}$.
As Mpc-scale outflows reach beyond the edges of groups, it was also necessary to estimate the pressure and baryon density in the AGN's more distant surroundings.
Following the bottom-right panel of \citet{Ricciardelli12013}'s Fig.~6, we set the baryon overdensity within voids at Porphyrion's redshift to $\delta = -0.7$.\footnote{In doing so, we implicitly assumed that the baryonic matter overdensity field is identical to the total matter overdensity field (which comprises contributions from both dark and baryonic matter), as \citet{Ricciardelli12013} considers the latter.}
We obtained a void baryon density $\rho_\mathrm{v} = \rho_\mathrm{c,0}\Omega_\mathrm{BM,0}(1+z)^3(1 + \delta) = 9 \cdot 10^{-31}\ \mathrm{g\ cm^{-3}}$, where $\rho_\mathrm{c,0}$ is today's critical density.
Following \citet{UptonSanderbeck12016}'s detailed study of IGM temperatures through cosmic time, which suggests a void temperature $T_\mathrm{v} \sim 10^{3}$--$10^{4}\ \mathrm{K}$ at Porphyrion's redshift, we set $T_\mathrm{v} = 1 \cdot 10^4\ \mathrm{K}$.
This choice reflects the fact that we are interested in void temperatures near the galaxy group.
Again applying the ideal gas law, and taking $\langle m \rangle$ as before, we obtained a void pressure $p_\mathrm{v} = 1 \cdot 10^{-19}\ \mathrm{Pa}$.
Finally, we defined the external pressure $p_\mathrm{e}(r) = p_\mathrm{g}(r) + p_\mathrm{v}$, baryon density $\rho_\mathrm{e}(r) = \rho_\mathrm{g}(r) + \rho_\mathrm{v}$, and baryon density--weighted temperature $T_\mathrm{e}(r) = \frac{\rho_\mathrm{g}(r)T_\mathrm{g} + \rho_\mathrm{v}T_\mathrm{v}}{\rho_\mathrm{e}(r)}$.
Figure~\ref{fig:dynamicalModel} shows these profiles.

We explored whether the addition of a filament component would significantly change Fig.~\ref{fig:dynamicalModel}'s profiles.
We assumed a baryon overdensity $\delta = 10$ at the filament spine, and baryon density and temperature profiles following \citet{Tuominen12021}'s results for massive filaments in the EAGLE simulation.
We found pressure and baryon density contributions of an importance similar to or lesser than that of the group, even at Mpc-scale distances.
We thus considered the addition of the filament unnecessary, especially in light of model uncertainties such as the group's mass and the surmised validity of extrapolating the group's UPP to Mpc-scale distances.

We generated 21 evolutionary tracks of 200 time steps each, spanning a range jet powers $Q = 10^{38.8}$--$10^{39.2}\ \mathrm{W}$.
Propagating total length and radio luminosity uncertainties, we obtained $Q = 1.3 \pm 0.1 \cdot 10^{39}\ \mathrm{W}$ and $T = 1.9\substack{+0.7\\-0.2}\ \mathrm{Gyr}$.
The outflow's jet power uncertainty is set by radio luminosity uncertainty while its age uncertainty is set by total length uncertainty.
Each jet's average speed $\langle\beta\rangle \coloneqq \frac{\langle v \rangle}{c} = \frac{l}{2cT} = 0.58\substack{+0.04\\-0.07}\%$, where $c$ is the speed of light.
The energy transported by the jets $E = QT = 7.6\substack{+2.1\\-0.7} \cdot 10^{55}\ \mathrm{J}$.
As a black hole can redirect at most half of the rest energy of infalling matter to electromagnetic radiation and jet fuelling, and the energy an RE AGN spends on electromagnetic radiation must at least equal the energy spent on jet fuelling, the black hole must have gained a mass $\Delta M_\bullet > 2\frac{E}{c^2} = 8.5\substack{+2.4\\-0.8} \cdot 10^8\ M_\odot$ while powering the jets.

\backmatter

\bmhead{Data availability}
The LoTSS DR2 is publicly available at \url{https://lofar-surveys.org/dr2_release.html}.
The authors will share the particular LOFAR, uGMRT, and Keck I Telescope data used in this work upon request.

\bmhead{Code availability}
The dynamical model used to interpret the outflow is described by \citet{Hardcastle12018} and available for download at \url{https://github.com/mhardcastle/analytic}.

\bmhead{Acknowledgments}
M.S.S.L.O. and R.J.v.W. acknowledge support from the VIDI research programme with project number 639.042.729, which is financed by the Dutch Research Council (NWO).
M.S.S.L.O. also acknowledges support from the CAS--NWO programme for radio astronomy with project number 629.001.024, which is financed by the NWO.
In addition, M.S.S.L.O., R.T., and R.J.v.W. acknowledge support from the ERC Starting Grant ClusterWeb 804208.
M.J.H. acknowledges support from the UK STFC [ST/V000624/1].
R.T. is grateful for support from the UKRI Future Leaders Fellowship (grant MR/T042842/1).
A.B. acknowledges financial support from the European Union - Next Generation EU.
The work of D.S. was carried out at the Jet Propulsion Laboratory, California Institute of Technology, under a contract with NASA.
We thank Frits Sweijen for making available \texttt{legacystamps} (\url{https://github.com/tikk3r/legacystamps}).
We thank Jesse van Oostrum and Riccardo Caniato for illuminating discussions.
LOFAR data products were provided by the LOFAR Surveys Key Science project (LSKSP; \url{https://lofar-surveys.org/}) and were derived from observations with the International LOFAR Telescope (ILT).
LOFAR \citep{vanHaarlem12013} is the Low Frequency Array designed and constructed by ASTRON. It has observing, data processing, and data storage facilities in several countries, which are owned by various parties (each with their own funding sources), and which are collectively operated by the ILT foundation under a joint scientific policy. The efforts of the LSKSP have benefited from funding from the European Research Council, NOVA, NWO, CNRS-INSU, the SURF Co-operative, the UK Science and Technology Funding Council, and the J\"ulich Supercomputing Centre.
We thank the staff of the GMRT that made these observations possible.
GMRT is run by the National Centre for Radio Astrophysics of the Tata Institute of Fundamental Research.
Some of the data presented herein were obtained at the W.\,M. Keck Observatory, which is operated as a scientific partnership among the California Institute of Technology, the University of California, and the National Aeronautics and Space Administration.
The Observatory was made possible by the generous financial support of the W.\,M. Keck Foundation.

\bmhead{Author contributions}
A.R.D.J.G.I.B.G. and M.S.S.L.O. discovered Porphyrion;
M.J.H., assisted by citizen scientists, independently found the outflow as part of LOFAR Galaxy Zoo.
M.S.S.L.O. coordinated the ensuing project.
R.J.v.W., H.J.A.R., and M.J.H. advised M.S.S.L.O. throughout.
A.B. and R.J.v.W. re-reduced and imaged the $6.2''$ LOFAR data.
R.T. reduced and imaged the $0.4''$ LOFAR data.
F.d.G. explored the use of LOFAR LBA data, which he reduced and imaged.
M.S.S.L.O. wrote the uGMRT follow-up proposal.
M.S.S.L.O. and H.T.I. reduced and imaged the uGMRT data.
S.G.D., D.S., and H.J.A.R. were instrumental in securing Keck time (P.I.: S.G.D.).
A.C.R. observed the host galaxy with LRIS; A.C.R. and D.S. reduced the data.
G.C.R. determined the host galaxy's SED and stellar mass; M.S.S.L.O. contributed.
M.J.H. performed dynamical modelling; M.S.S.L.O. contributed.
M.S.S.L.O. derived the deprojection, void penetration probability, filament heating, metallicity, and diffusion and magnetogenesis formulae.
M.S.S.L.O. wrote the article, with contributions from A.R.D.J.G.I.B.G., R.T., and A.C.R.
All authors provided comments to improve the text.

\bmhead{Competing interests}
The authors declare no competing interests.

\footnotesize
\bibliography{sn-article}%

\clearpage

\begin{appendices}
\normalsize
\section{Supplementary Methods}

\paragraph{Total outflow length}
To estimate Porphyrion's total length from its projected length, we perform statistical deprojection.
Equation~9 of \citet{Oei12023} stipulates the probability density function (PDF) of an outflow's total length random variable (RV) $L$ in case its projected length RV $L_\mathrm{p}$ is known to equal some value $l_\mathrm{p}$.
This PDF is parametrised by the tail index $\xi$ of the Pareto distribution assumed to describe $L$.
We calculate the median and expectation value of $L\ \vert\ L_\mathrm{p} = l_\mathrm{p}$ for tail indices $\xi = -3$ and $\xi = -4$, the integer values closest to the observationally favoured $\xi = -3.5 \pm 0.5$ \citep{Oei12023}.

First, we determine the cumulative distribution function (CDF) of $L\ \vert\ L_\mathrm{p} = l_\mathrm{p}$ through integration:
\begin{align}
F_{L \vert L_\mathrm{p} = l_\mathrm{p}}(l) &\coloneqq \int_{-\infty}^l f_{L \vert L_\mathrm{p} = l_\mathrm{p}}(l')\ \mathrm{d}l'\\
&=\frac{-\xi}{2^{1+\xi}\pi}\frac{\Gamma^2\left(-\frac{\xi}{2}\right)}{\Gamma(-\xi)} \int_{1}^{\max{\{x,1\}}} \frac{x'^{\xi-1}}{\sqrt{x'^2-1}}\ \mathrm{d}x',\nonumber
\end{align}
where $x \coloneqq \frac{l}{l_\mathrm{p}}$ and $x' \coloneqq \frac{l'}{l_\mathrm{p}}$.

For $\xi = -3$, the CDF concretises to
\begin{align}
    F_{L \vert L_\mathrm{p} = l_\mathrm{p}}(l) &= \frac{3}{2} \int_{1}^{\max{\{x,1\}}} \frac{\mathrm{d}x'}{x'^4\sqrt{x'^2-1}}\\
&=\begin{cases}
0 & \text{if } x < 1;\\
\frac{(2x^2+1)\sqrt{x^2-1}}{2 x^3} & \text{if } x \geq 1.
\end{cases}\nonumber
\end{align}
The median conditional total length, $l_\mathrm{m}$, is defined by $F_{L \vert L_\mathrm{p} = l_\mathrm{p}}(l_\mathrm{m}) \coloneqq \frac{1}{2}$.
Numerically, we obtain $x_\mathrm{m} \coloneqq \frac{l_\mathrm{m}}{l_\mathrm{p}} \approx 1.0664$, or $l_\mathrm{m} \approx 1.0664\ l_\mathrm{p}$.
As $l_\mathrm{p} = 6.43 \pm 0.05\ \mathrm{Mpc}$, we find $l_\mathrm{m} = 6.86 \pm 0.05\ \mathrm{Mpc}$.
An analogous numerical determination of the 16-th and 84-th percentiles then yields $l = 6.9\substack{+1.6\\-0.4}\ \mathrm{Mpc}$.

For $\xi = -4$, the CDF concretises to
\begin{align}
    F_{L \vert L_\mathrm{p} = l_\mathrm{p}}(l) &= \frac{16}{3\pi} \int_{1}^{\max{\{x,1\}}} \frac{\mathrm{d}x'}{x'^5\sqrt{x'^2-1}}\\
    &=\begin{cases}
        0 & \text{if } x < 1;\\
        \frac{2}{3\pi}\left(\frac{(3 x^2 + 2)\sqrt{x^2-1}}{x^4}+ 3\arccos{\frac{1}{x}}\right) & \text{if } x \geq 1.
    \end{cases}\nonumber
\end{align}
Numerically, we obtain $x_\mathrm{m} \approx 1.0515$, or $l_\mathrm{m} \approx 1.0515\ l_\mathrm{p}$, and thus $l_\mathrm{m} = 6.76 \pm 0.05\ \mathrm{Mpc}$.
In the same way as before, we find $l = 6.8\substack{+1.2\\-0.3}\ \mathrm{Mpc}$.

Equation~10 of \citet{Oei12023} gives a closed-form expression for $\mathbb{E}\left[L\ \vert\ L_\mathrm{p} = l_\mathrm{p}\right](\xi)$.
Table~1 of the same work lists $\mathbb{E}[L\ \vert\ L_\mathrm{p} = l_\mathrm{p}](\xi = -3) = \frac{3\pi}{8} l_\mathrm{p}$ and $\mathbb{E}[L\ \vert\ L_\mathrm{p} = l_\mathrm{p}](\xi = -4) = \frac{32}{9\pi} l_\mathrm{p}$.
In the case of Porphyrion, these expressions concretise to $\mathbb{E}[L\ \vert\ L_\mathrm{p} = l_\mathrm{p}](\xi = -3) = 7.58 \pm 0.06\ \mathrm{Mpc}$ and $\mathbb{E}[L\ \vert\ L_\mathrm{p} = l_\mathrm{p}](\xi = -4) = 7.28 \pm 0.05\ \mathrm{Mpc}$.

By conditioning $L$ on more knowledge than a value for $L_\mathrm{p}$ alone, statistical deprojection could be made more precise.
For example, one could additionally condition on the fact that Porphyrion is generated by a Type 2 radiatively efficient (RE) AGN.
If Type 1 RE AGN are seen mostly face-on and Type 2 RE AGN are seen mostly edge-on, as proposed by the unification model \citep[e.g.][]{Heckman12014}, then the detection of a Type 2 RE AGN would imply that the jets make a small angle with the sky plane.
Extending the formulae to include this knowledge is beyond the scope of this work; however, mindful of the associated deprojection factor--reducing effect, we choose $\xi = -4$ as our fiducial tail index.

To assess Porphyrion's transport capabilities in a cosmological context, it is instructive to calculate its length relative to Cosmic Web length scales.
In particular, the outflow's total length relative to the typical cosmic void radius at its epoch is $f_\mathrm{v} \coloneqq l(1+z)R_\mathrm{v}^{-1}$, where $R_\mathrm{v}$ is the typical comoving cosmic void radius.
For $l = 6.8\substack{+1.2\\-0.3}\ \mathrm{Mpc}$, $z = 0.896 \pm 0.001$, and $R_\mathrm{v} = 20\ \mathrm{Mpc}$ \citep{Correa12021}, we find $f_\mathrm{v} = 64\substack{+12\\-2}\ \%$.
For our fiducial total length $l = 7\ \mathrm{Mpc}$, we find $f_\mathrm{v} = 66\%$.

\paragraph{Void penetration probability}
Porphyrion's orientation relative to its native Cosmic Web filament is currently unknown.
We calculate the probability that an outflow breaches its filament, thus penetrating the surrounding voids, by assuming that jet orientations are independent from filament orientations.
We furthermore assume that the jets are straight and of equal length, that the filament is of cylindrical shape, and that the host galaxy resides at the filament's spine, where the gravitational potential is lowest.
The RV $\Theta_\mathrm{f}$ denotes the angle between the jet axis and the filament axis, whilst the constants $R_\mathrm{f}$ and $D_\mathrm{f} \coloneqq 2 R_\mathrm{f}$ denote the filament radius and diameter, respectively.
An outflow of total length $l$ penetrates voids with probability
\begin{align}
    \mathbb{P}\left(L \sin{\Theta_\mathrm{f}} > D_\mathrm{f}\ \vert\ L = l\right) &= \mathbb{P}\left(\sin{\Theta_\mathrm{f}} > \frac{D_\mathrm{f}}{l}\right)\nonumber\\
    &= 1 - F_{\sin{\Theta_\mathrm{f}}}\left(\frac{D_\mathrm{f}}{l}\right).
\label{eq:voidPenetrationProbabilityLengthTotal}
\end{align}
The RV $\Theta_\mathrm{f}$ has support on the interval $0 < \theta \leq \frac{\pi}{2}$.
On this interval, the CDF is
\begin{align}
F_{\Theta_\mathrm{f}}(\theta) \coloneqq \mathbb{P}(\Theta_\mathrm{f} \leq \theta) &= \frac{1}{2\pi}\int_0^{\theta} \int_0^{2\pi} \sin{\theta'}\ \mathrm{d}\varphi\ \mathrm{d}\theta'\nonumber\\
&= 1 - \cos{\theta}.
\end{align}
The RV $\sin{\Theta_\mathrm{f}}$ has support on the interval $0 < y \leq 1$.
On this interval, the CDF is
\begin{align}
    F_{\sin{\Theta_\mathrm{f}}}(y) &\coloneqq \mathbb{P}\left(\sin{\Theta_\mathrm{f}} \leq y\right) = \mathbb{P}\left(\Theta_\mathrm{f} \leq \arcsin{y}\right)\nonumber\\
    &= 1 - \cos{\arcsin{y}} = 1 - \sqrt{1 - y^2},
\label{eq:CDFSineAngle}
\end{align}
where we use the fact that the arcsine function is monotonically increasing.
By combining Eqs.~\ref{eq:voidPenetrationProbabilityLengthTotal} and \ref{eq:CDFSineAngle}, we obtain
\begin{align}
    \mathbb{P}\left(L \sin{\Theta_\mathrm{f}} > D_\mathrm{f}\ \vert\ L = l\right) = \begin{cases}
    0 & \text{if } l \leq D_\mathrm{f};\\
    \sqrt{1 - \left(\frac{D_\mathrm{f}}{l}\right)^2} & \text{if } l > D_\mathrm{f}.
    \end{cases}
\end{align}
Typically, however, we only know an outflow's \emph{projected} length --- not its \emph{total} length.
The quantity of highest practical interest therefore is
\begin{align}
p_\mathrm{v} \coloneqq&\ \mathbb{P}\left(L \sin{\Theta_\mathrm{f}} > D_\mathrm{f}\ \vert\ L_\mathrm{p} = l_\mathrm{p}\right)\nonumber\\
=&\ \mathbb{P}\left(L_\mathrm{p} \frac{\sin{\Theta_\mathrm{f}}}{\sin{\Theta_\mathrm{o}}} > D_\mathrm{f}\ \vert\ L_\mathrm{p} = l_\mathrm{p}\right)\nonumber\\
=&\ \mathbb{P}\left(\frac{\sin{\Theta_\mathrm{f}}}{\sin{\Theta_\mathrm{o}}} > \frac{D_\mathrm{f}}{l_\mathrm{p}}\right) = 1 - F_X\left(\frac{D_\mathrm{f}}{l_\mathrm{p}}\right),
\end{align}
where the RV $\Theta_\mathrm{o}$ is the angle between the jet axis and the line of sight, and the RV $X$ is the ratio of the RVs $\sin{\Theta_\mathrm{f}}$ and $\sin{\Theta_\mathrm{o}}$.
These latter RVs are independent and identically distributed: $F_{\sin{\Theta_\mathrm{f}}}(y) = F_{\sin{\Theta_\mathrm{o}}}(y)$.
We derive the PDF $f_X$ by using the standard formula for the ratio distribution PDF (for independent RVs).
This formula demands the determination of $f_{\sin{\Theta_\mathrm{f}}}(y) = f_{\sin{\Theta_\mathrm{o}}}(y)$.
From Eq.~\ref{eq:CDFSineAngle}, we find that on the interval $0 < y \leq 1$, the PDF is
\begin{align}
    f_{\sin{\Theta_\mathrm{f}}}(y) = \frac{\mathrm{d}}{\mathrm{d}y}F_{\sin{\Theta_\mathrm{f}}}(y) = \frac{y}{\sqrt{1-y^2}}.
\end{align}
To find the distribution of $X$, it is helpful to distinguish three intervals.
For $x \leq 0$, $f_X(x) = 0$, because $X$ is the ratio of two positive RVs.
Then, for $0 < x < 1$,
\begin{align}
    f_X(x) = x \int_0^1 \frac{y^3}{\sqrt{1-x^2y^2}\sqrt{1-y^2}}\ \mathrm{d}y,
\end{align}
while for $x > 1$,
\begin{align}
    f_X(x) = x \int_0^{\frac{1}{x}} \frac{y^3}{\sqrt{1-x^2y^2}\sqrt{1-y^2}}\ \mathrm{d}y.
\end{align}
Solving the integrals leads to
\begin{align}
    f_X(x) = \begin{cases}
    0 & \text{if } x \leq 0;\\
    \frac{(x^2+1) \ln{\frac{1+x}{1-x}} - 2x}{4x^2} & \text{if } 0 < x < 1;\\
    \frac{(x^2+1) \ln{\frac{x + 1}{x - 1}} - 2x}{4x^2} & \text{if } x > 1.
    \end{cases}
\end{align}
At $x = 1$, $f_X$ is undefined.
\begin{figure}[t]
    \centering
    \includegraphics[width=\columnwidth]{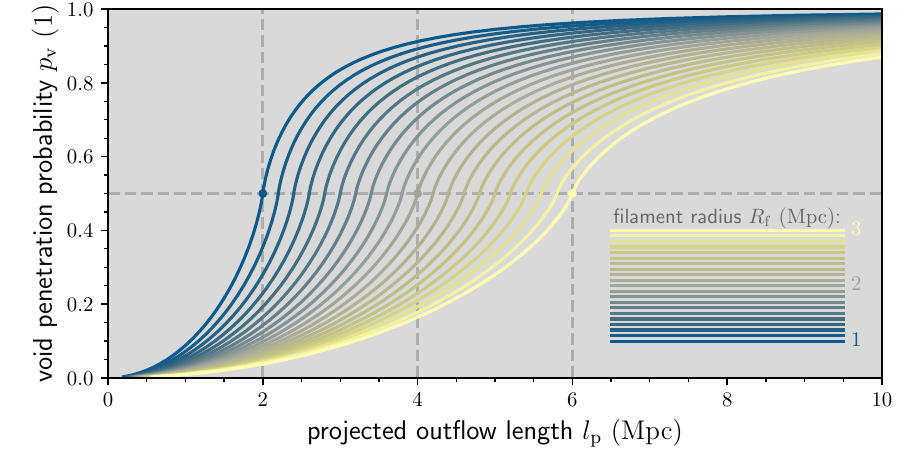}
    \caption{
    \textbf{
    From its projected length and an estimated radius for its native filament, one can calculate the probability that an outflow penetrates the surrounding voids.
    }
    Outflows with projected lengths equalling the diameters of their filaments penetrate voids with $50\%$ probability (as the dots exemplify).
    }
    \label{fig:probabilityVoidPenetration}
\end{figure}\noindent
We calculate $F_X$ through integration, yielding
\begin{align}
&F_X(x) =\begin{cases}
0 & \text{if } x \leq 0;\\
\frac{1}{2} + \frac{1}{4}\left(\left(x + 1 - \frac{1}{x}\right) \ln{\frac{1+x}{1-x}}+\ln{\frac{(1-x)^2}{1-x^2}}\right) & \text{if } 0 < x < 1;\\
\frac{1}{2} + \frac{1}{4}\frac{x^2-1}{x}\ln{\frac{x+1}{x-1}} & \text{if } x > 1.
\end{cases}
\end{align}
$F_X$ is continuously extendable at $x = 1$ by defining $F_X(1) \coloneqq \frac{1}{2}$.
Finally, we define $x \coloneqq \frac{D_\mathrm{f}}{l_\mathrm{p}}$, so that $p_\mathrm{v}(x) = 1 - F_X(x)$:
\begin{align}
    p_\mathrm{v}(x) = \begin{cases}
\frac{1}{2} - \frac{1}{4}\left(\left(x + 1 - \frac{1}{x}\right) \ln{\frac{1+x}{1-x}} + \ln{\frac{(1-x)^2}{1-x^2}}\right) & \text{if } 0 < x < 1;\\
\frac{1}{2} & \text{if } x = 1;\\
\frac{1}{2} - \frac{1}{4}\frac{x^2-1}{x}\ln{\frac{x+1}{x-1}} & \text{if } x > 1.
\end{cases}
\end{align}
The median of $X$ equals unity.
Therefore, half of the outflows with projected lengths $l_\mathrm{p} = D_\mathrm{f}$ penetrate voids.
For outflows with larger projected lengths, void penetration is more likely than not.
Figure~\ref{fig:probabilityVoidPenetration} shows void penetration probabilities for physically relevant parameter ranges.

For Porphyrion, we take $l_\mathrm{p}$ as before and adopt a filament (core) radius $R_\mathrm{f} = 1.2\ \mathrm{Mpc}$ \citep{Tuominen12021}; this yields $x = 0.373 \pm 0.003$ and therefore $p_\mathrm{v} = 95.2 \pm 0.1\%$.

\paragraph{Filament shape modification}
We predict that powerful, long-lived outflows like Porphyrion cause their host galaxies' filaments to expand thermally.
Through lateral shocks, the jets distribute an amount of heat $Q$ over the warm--hot IGM.
This medium is sufficiently dilute that plasma interactions can be neglected; as a result, the ideal gas law, $pV = N k_\mathrm{B} T$, may be adopted as the equation of state.
Here, $p$, $V$, $N$, and $T$ are the filament's pressure, volume, plasma particle number, and temperature, respectively; $k_\mathrm{B}$ is Boltzmann's constant.
Assuming a thermodynamic process at constant pressure and particle number, the work $W$ is
\begin{align}
    W = p \Delta V = N k_\mathrm{B} \Delta T.
\label{eq:work}
\end{align}
Before the outflow's emergence, the filament's equation of state is $p V_\mathrm{i} = N k_\mathrm{B} T_\mathrm{i}$, where $V_\mathrm{i}$ and $T_\mathrm{i}$ are its initial volume and temperature, respectively.
Upon dividing Eq.~\ref{eq:work} by this equation of state, one obtains
\begin{align}
    \frac{\Delta V}{V_\mathrm{i}} = \frac{\Delta T}{T_\mathrm{i}}.
\end{align}
Assuming that the filament retains a cylindrical shape, initially with radius $r_\mathrm{i}$ and finally with radius $r_\mathrm{f}$, and using that $\Delta V \coloneqq V_\mathrm{f} - V_\mathrm{i}$, one obtains
\begin{align}
    \frac{r_\mathrm{f}}{r_\mathrm{i}} = \sqrt{1 + \frac{\Delta T}{T_\mathrm{i}}}.
\label{eq:radiusRatio}
\end{align}
The radius ratio, $\frac{r_\mathrm{f}}{r_\mathrm{i}}$, depends only on the ratio between the temperature increase $\Delta T \coloneqq T_\mathrm{f} - T_\mathrm{i}$ and the initial temperature.
The temperature increase is
\begin{align}
    \Delta T = \frac{Q}{N C_\mathrm{p,m}},
\end{align}
where $C_\mathrm{p,m}$ is the molar heat capacity at constant pressure.
For a monatomic gas or a hydrogen plasma, $C_\mathrm{p,m} = \frac{5}{2}R$, where $R$ is the molar gas constant.
The number of filamentary electrons and atomic nuclei affected by the outflow is
\begin{align}
    N = \frac{\pi r_\mathrm{i}^2 L \rho_\mathrm{i}}{\mu m_\mathrm{p}},
\end{align}
where $L$ is the length of the cylindrical segment affected, $\rho_\mathrm{i}$ is the initial baryonic mass density, $\mu$ is the average mass of a plasma particle relative to the proton mass, and $m_\mathrm{p}$ is the proton mass.
We estimate $\frac{L}{2}$ by multiplying the typical speed of lateral shocks with the outflow's lifetime. 
We decompose $\rho_\mathrm{i} = \rho_\mathrm{c,0} \Omega_\mathrm{BM,0} (1+z)^3 (1 + \delta)$, where $z$ and $\delta$ are the filament's cosmological redshift and baryonic overdensity, respectively.

We assess the outflow-induced morphological change to Porphyrion's filament by evaluating Eq.~\ref{eq:radiusRatio}, taking $Q = 10^{55}\ \mathrm{J}$, $r_\mathrm{i} = 1.2\ \mathrm{Mpc}$, $L = 7\ \mathrm{Mpc}$, $z = 0.9$, $1 + \delta = 10$, $\mu = 0.5$, and $T_\mathrm{i} = 10^7\ \mathrm{K}$; we find $\Delta T = 9 \cdot 10^6\ \mathrm{K}$ and $r_\mathrm{f} = 1.7\ \mathrm{Mpc}$ (an increase of ${\sim}40\%$).
Porphyrion's heat dissipation renders the outflow's native filament much hotter and thicker than it would have otherwise been.

\paragraph{Transport of heavy elements to voids}
Outflows from RE AGN could contain more heavy atoms than outflows from RI AGN: RE AGN tend to reside in galaxies with higher star formation rates and thus more vigorous stellar winds, suggesting increased entrainment of wind-borne atomic nuclei into jets \citep{Wykes12015}.
The order-of-magnitude calculations presented here, to be verified by future simulations, indicate that Mpc-scale outflows could supply $\mathrm{Mpc}^3$-scale volumes in cosmic voids with significant heavy atoms, though consistent with upper limits \citep{Stocke12007}.

We calculated the heavy element enrichment of the IGM in voids due to the deposition of atomic nuclei initially entrained in the jets of Mpc-scale outflows.
In particular, we estimated the final metallicity in the deposition region considering both internal and external entrainment.
Internal entrainment refers to the entrainment into jets of atomic nuclei from stellar winds, \emph{internal} to the host galaxy.
External entrainment refers to the entrainment into jets of atomic nuclei dwelling in the IGM, \emph{external} to the host galaxy.

Denoting the (two-sided) internal mass entrainment rate by $\dot{M}_\mathrm{i}$, which we assumed constant through time, the total internally entrained mass by an outflow of age $T$ is
\begin{align}
    M_\mathrm{i} = \dot{M}_\mathrm{i}T.
\end{align}
The internal mass entrainment rates of Centaurus A and 3C 31, both Fanaroff--Riley I (FR I) outflows in the Local Universe, are estimated to be $\dot{M}_\mathrm{i} = 2 \cdot 10^{-3}\ M_\odot\ \mathrm{yr}^{-1}$ \citep{Wykes12015} and $\dot{M}_\mathrm{i} = 8 \cdot 10^{-3}\ M_\odot\ \mathrm{yr}^{-1}$ \citep{Laing12002}, respectively.
At $M_\star = 6.3 \cdot 10^{11}\ M_\odot$, 3C 31's host stellar mass is similar to Porphyrion's.
However, Porphyrion is a Fanaroff--Riley II (FR II) outflow, suggesting a smaller jet opening angle $\omega$ and thus a \emph{smaller} $\dot{M}_\mathrm{i}$.
If $\omega$ is a factor of order unity smaller for FR II outflows than for FR I outflows \citep[e.g.][]{Pushkarev12009, Laing12014}, and $\dot{M}_\mathrm{i} \propto \omega^2$, then $\dot{M}_\mathrm{FR\,II} \sim 10^{-1} \dot{M}_\mathrm{FR\,I}$.
On the other hand, Porphyrion's host is seen much closer to the cosmic heyday of star formation, suggesting a larger SFR and thus a \emph{larger} $\dot{M}_\mathrm{i}$.
After $z = 1.9$, SFRs $S$ typically decayed exponentially with an $e$-folding time of $3.9\ \mathrm{Gyr}$ \citep[e.g.][]{Madau12014}.
If $\dot{M}_\mathrm{i} \propto S$, then $\dot{M}_\mathrm{i}(z=z_\mathrm{P}) \sim e^2 \dot{M}_\mathrm{i}(z=0) \sim 10^1 \dot{M}_\mathrm{i}(z=0)$, where $z_\mathrm{P}$ is Porphyrion's redshift.
Assuming that both effects are indeed of comparable importance, we provisionally adopted 3C 31's $\dot{M}_\mathrm{i} = 8 \cdot 10^{-3}\ M_\odot\ \mathrm{yr}^{-1}$ as our fiducial value.
Taking $T$ as before, we obtained $M_\mathrm{i} = 2 \cdot 10^7\ M_\odot$.
The total externally entrained mass is
\begin{align}
    M_\mathrm{e} = 2 \int_0^{\frac{l}{2}} \rho_\mathrm{e}(r) A_\mathrm{e}(r)\ \mathrm{d}r,
\end{align}
where $A_\mathrm{e}(r)$ is the entrainment cross-section at a distance $r$ from the AGN.
Perhaps the simplest approach is to parametrise $A_\mathrm{e} \eqqcolon \pi R_\mathrm{e}^2$, where $R_\mathrm{e}$ is a (constant) effective radius defined such that all baryons closer to the jet axis than $R_\mathrm{e}$ are entrained.
Taking $l$ and $\rho_\mathrm{e}$ as before, and $R_\mathrm{e} = 1\ \mathrm{kpc}$ \citep[e.g.][]{Perucho12023}, we obtained $M_\mathrm{e} = 4 \cdot 10^7\ M_\odot$.
Although highly uncertain, these estimates suggest that $M_\mathrm{i}$ and $M_\mathrm{e}$ can be of the same order of magnitude.

The total internally entrained mass in heavy elements is $Z_\mathrm{i}M_\mathrm{i}$, where $Z_\mathrm{i}$ is the mass-weighted mean metallicity of the galaxy's stellar winds.
The total externally entrained mass in heavy elements is $Z_\mathrm{e}M_\mathrm{e}$, where $Z_\mathrm{e}$ is the mass-weighted metallicity of the IGM along the jet.
Assuming that the IGM in the voids is initially pristine, its final metallicity is
\begin{align}
    Z_\mathrm{v} = \frac{Z_\mathrm{i}M_\mathrm{i} + Z_\mathrm{e} M_\mathrm{e}}{M_\mathrm{i} + M_\mathrm{e} + \rho_\mathrm{v} V},
\end{align}
where $\rho_\mathrm{v}$ is the baryon mass density within a deposition region of volume $V$.
Taking a spherical deposition region with a diameter of $1\ \mathrm{Mpc}$, and $\rho_\mathrm{v}$ as before, we obtained $\rho_\mathrm{v} V = 7 \cdot 10^{9}\ M_\odot$.
Assuming $Z_\mathrm{i} = Z_\odot$ \citep[e.g.][]{Wykes12015} and $Z_\mathrm{e} = 10^{-1}\ Z_\odot$ \citep[e.g.][]{Mernier12018}, we found $Z_\mathrm{v} = 3 \cdot 10^{-3}\ Z_\odot$.
In conclusion, order-of-magnitude arguments suggest that void-penetrating Mpc-scale outflows can endow the local IGM with metallicities $Z_\mathrm{v} \sim 10^{-3}$--$10^{-2}\ Z_\odot$.

\paragraph{Quasar mass--based host galaxy candidate elimination}
SDSS J152933.03+601552.5 is the quasar-hosting galaxy $19''$ north-northeast of J152932.16+601534.4, the galaxy we have identified as Porphyrion's host.
We initially also considered SDSS J152933.03+601552.5 as a host galaxy candidate.
However, aforementioned arguments involving the presence of jets and their orientation and, to a lesser degree, arguments involving core radio luminosity and core synchrotron self-absorption all favour J152932.16+601534.4.
We now discuss how our results would change if, instead, SDSS J152933.03+601552.5 were Porphyrion's host galaxy.
Doing so will lead to a contradiction that disproves this alternative hypothesis.

First, we discuss results that do not require dynamical modelling.
To start with, Porphyrion would remain generated by an RE AGN.
The host galaxy redshift would decrease from $z = 0.896 \pm 0.001$ to $z = 0.799 \pm 0.001$, decreasing Porphyrion's projected length from $l_\mathrm{p} = 6.43 \pm 0.05\ \mathrm{Mpc}$ to $l_\mathrm{p} = 6.21 \pm 0.05\ \mathrm{Mpc}$.
Again using $\xi = -4$, the total length would decrease from $l = 6.8\substack{+1.2\\-0.3}\ \mathrm{Mpc}$ to $l = 6.5\substack{+1.2\\-0.3}\ \mathrm{Mpc}$ and its conditional expectation from $\mathbb{E}[L\ \vert\ L_\mathrm{p} = l_\mathrm{p}] = 7.28 \pm 0.05\ \mathrm{Mpc}$ to $\mathbb{E}[L\ \vert\ L_\mathrm{p} = l_\mathrm{p}] = 7.03 \pm 0.06\ \mathrm{Mpc}$.
If orientation distinguishes Type 1 from Type 2 RE AGN, as the unification model supposes, then these statistical deprojection results may underestimate Porphyrion's total length.
Porphyrion would remain the projectively largest galaxy-made structure identified so far.
Porphyrion's total radio luminosity at rest-frame wavelength $\lambda_\mathrm{r} = 2\ \mathrm{m}$ would decrease from $L_\nu = 2.8 \pm 0.3\ \cdot 10^{26}\ \mathrm{W\ Hz^{-1}}$ to $L_\nu = 2.2 \pm 0.2\ \cdot 10^{26}\ \mathrm{W\ Hz^{-1}}$.

Next, we discuss results that come from dynamical modelling.
The jet power would decrease from $Q = 1.3 \pm 0.1 \cdot 10^{39}\ \mathrm{W}$ to $Q = 1.0 \pm 0.1 \cdot 10^{39}\ \mathrm{W}$, while the age would slightly increase from $T = 1.9\substack{+0.7\\-0.2}\ \mathrm{Gyr}$ to $T = 1.9\substack{+0.7\\-0.1}\ \mathrm{Gyr}$.\footnote{Significant jet-mediated transport of heavy elements to the IGM would remain plausible.
The host's stellar mass would decrease from $M_\star = 6.7 \pm 1.4 \cdot 10^{11}\ M_\odot$ to $M_\star = 4.0\substack{+0.3\\-0.3} \cdot 10^{11}\ M_\odot$, while the SFR would become $S = 4.9\substack{+0.3\\-0.4} \cdot 10^1\ M_\odot\ \mathrm{yr}^{-1}$ \citep{Barrows12021}.
}
The transported energy would decrease from $E = 7.6\substack{+2.1\\-0.7} \cdot 10^{55}\ \mathrm{J}$ to $E = 6.4\substack{+1.8\\-0.6} \cdot 10^{55}\ \mathrm{J}$, and the black hole mass gain from $\Delta M_\bullet > 8.5\substack{+2.4\\-0.8} \cdot 10^8\ M_\odot$ to $\Delta M_\bullet > 7.2\substack{+2.0\\-0.7} \cdot 10^8\ M_\odot$.

Finally, we arrive at a contradiction, as the quasar's SMBH mass (measured from its SDSS BOSS spectrum) $M_\bullet = 2.5 \pm 0.3 \cdot 10^8\ M_\odot$ \citep{Chen12018}.
This mass is lower than the minimum mass gain associated to the fuelling of Porphyrion's jets.
Thus, assuming that SDSS J152933.03+601552.5 is the outflow's host galaxy leads to a contradiction.
This argument reaffirms that J152932.16+601534.4 is Porphyrion's host.

\paragraph{Diffusion of lobe plasma through voids}
When cosmic rays move through the jumbled magnetic fields of galaxy clusters and filaments of the Cosmic Web, the Lorentz force scatters them repeatedly.
The mean free path of the ensuing random walk is so short that the CRs radiate away their energy before they are able to travel a cosmologically significant distance \citep[e.g.][]{Brunetti12014}.
Clusters and filaments thus effectively lock into place the CRs that are injected into them.
By contrast, magnetic fields with Mpc-scale coherence lengths in voids are orders of magnitude weaker than those in clusters and filaments \citep[e.g.][]{Chen12015}, and as a result, CRs that are released into voids might diffuse through their entirety within a few gigayears.
Void-filling diffusion of CRs might be especially rapid at early epochs: during Porphyrion's lifetime, for instance, the proper volumes of voids were on average an order of magnitude smaller than they are today.

Consider a void region filled with relativistic particles, so that their velocity components obey
\begin{align}
    v_x^2 + v_y^2 + v_z^2 \approx c^2.
\label{eq:velocityComponentsRelativisticParticle}
\end{align}
We treat $v_x$, $v_y$, and $v_z$ as random variables subject to the above constraint.
If the particles have no bulk motion, and move in all directions with equal probability density,
\begin{align}
\mathbb{E}[v_x^N] = \mathbb{E}[v_y^N] = \mathbb{E}[v_z^N]
\label{eq:velocityComponentsIsotropy}
\end{align}
for any $N \in \mathbb{R}$.
In particular, given the absence of bulk motion, $\mathbb{E}[v_x] = \mathbb{E}[v_y] = \mathbb{E}[v_z] = 0$.
By taking expectations on both sides of Eq.~\ref{eq:velocityComponentsRelativisticParticle}, using the linearity of expectation, and invoking Eq.~\ref{eq:velocityComponentsIsotropy}, we find
\begin{align}
    \mathbb{E}[v_x^2] = \mathbb{E}[v_y^2] = \mathbb{E}[v_z^2] = \frac{c^2}{3}.
\label{eq:velocityComponentsSquaredExpectation}
\end{align}
Without loss of generality, we assume the region's magnetic field $\vec{B}$ to be oriented along the $z$-axis.
The speed perpendicular to $\vec{B}$ is $v_\perp = \sqrt{v_x^2 + v_y^2}$, so that, upon invoking Eq.~\ref{eq:velocityComponentsSquaredExpectation}, we find $\mathbb{E}[v_\perp^2] = \frac{2}{3}c^2$.
A typical speed for relativistic particles perpendicular to a magnetic field thus is
\begin{align}
    \sqrt{\mathbb{E}[v_\perp^2]} = \sqrt{\frac{2}{3}}c \approx 0.8165\ c.
\label{eq:velocityPerpendicularTypical}
\end{align}
Starting from Fick's first law of diffusion, and solving the case of Brownian motion in three dimensions, one obtains
\begin{align}
    r = \sqrt{6 D t},
\end{align}
where $r$ is the typical proper distance to the particles' origin after a time $t$.
To find the diffusion coefficient $D$, we consider Bohm diffusion, in which charged particles diffuse through a turbulent magnetic field as a result of the Lorentz force.
Whereas predicting the trajectory of any single charged particle requires knowledge of the specific magnetic field structure in its surroundings, the statistical properties of Bohm diffusion are determined solely by the statistical properties of the magnetic field and the charge and energy of the diffusing particles.
The Larmor radius for a particle with Lorentz factor $\gamma$, total velocity $v$, rest mass $m$, and charge $q$, is
\begin{align}
    r_\mathrm{L} = \frac{\gamma(v)m v_\perp}{|q|B}.
\end{align}
For a relativistic particle whose $v_\perp$ is given by Eq.~\ref{eq:velocityPerpendicularTypical}, we obtain a Larmor radius
\begin{align}
    r_\mathrm{L}(E) = \sqrt{\frac{2}{3}}\frac{E}{c|q|B} = 8.8 \cdot 10^2\ \mathrm{pc} \cdot \frac{E}{1\ \mathrm{GeV}} \cdot \frac{10^{-15}\ \mathrm{G}}{B},
\end{align}
where $E$ is the total (i.e. rest plus kinetic) energy of the particle.
The diffusion coefficient for charged particles in a magnetic field with a Kolmogorov turbulence spectrum is well approximated \citep{Globus12008} by
\begin{align}
    D(E) \approx D_\mathrm{Bohm}(E_0) \left(\frac{E}{E_0}\right)^{\frac{1}{3}} + D_\mathrm{Bohm}(E_1) \left(\frac{E}{E_1}\right)^2.
\end{align}
Here, $E_0$ is the energy for which the circumference of gyration equals the magnetic field coherence length $\lambda_\mathrm{c}$:
\begin{align}
    2 \pi r_\mathrm{L}(E_0) = \lambda_\mathrm{c}.
\end{align}
For $q = \pm e$, where $e$ is the elementary charge, this equation implies that
\begin{align}
    E_0 = \sqrt{\frac{3}{2}}\frac{\lambda_\mathrm{c} c|q|B}{2\pi} = 1.8 \cdot 10^2\ \mathrm{GeV} \cdot \frac{\lambda_\mathrm{c}}{1\ \mathrm{Mpc}} \cdot \frac{B}{10^{-15}\ \mathrm{G}}.
\end{align}
Furthermore, $E_1 = \frac{3}{2}E_0$.
The Bohm diffusion coefficient \citep[e.g.][]{Globus12008} $D_\mathrm{Bohm}$ is
\begin{align}
    D_\mathrm{Bohm}(E) &= \frac{c}{3}r_\mathrm{L}(E)\\
    &= 9.0 \cdot 10^{-2}\ \frac{\mathrm{Mpc}^2}{\mathrm{Gyr}} \cdot \frac{E}{1\ \mathrm{GeV}} \cdot \frac{10^{-15}\ \mathrm{G}}{B}.
\end{align}
We note that
\begin{align}
    D_\mathrm{Bohm}(E_0) = \frac{c}{6\pi}\lambda_\mathrm{c} = 1.6 \cdot 10^1 \frac{\mathrm{Mpc}^2}{\mathrm{Gyr}} \cdot \frac{\lambda_\mathrm{c}}{1\ \mathrm{Mpc}}
\end{align}
is independent of the void's magnetic field strength.
Because $D_\mathrm{Bohm} \propto r_\mathrm{L} \propto E$, we have $D_\mathrm{Bohm}(E_1) = \frac{3}{2} D_\mathrm{Bohm}(E_0)$.

For $E = 1\ \mathrm{GeV}$, $\lambda_\mathrm{c} = 1\ \mathrm{Mpc}$, and $B = 10^{-15}\ \mathrm{G}$, we find $D(E) = 2.9\ \frac{\mathrm{Mpc}^2}{\mathrm{Gyr}}$.
After $t = 1\ \mathrm{Gyr}$, the typical displacement of cosmic rays that escaped from the outflow's lobes is $r = 4.2\ \mathrm{Mpc}$.
We note that $r$ scales slowly with particle energy, magnetic field strength, and (to a lesser degree) with coherence length:
\begin{align}
    r \propto E^{\frac{1}{6}} B^{-\frac{1}{6}} \lambda_\mathrm{c}^{\frac{1}{3}} t^{\frac{1}{2}}.
\end{align}
For short time intervals $t$, we can ignore the expansion of the Universe; defining $r_\mathrm{c} \coloneqq r(1+z)$, the void volume-filling fraction $\mathcal{V}$ of a single lobe becomes
\begin{align}
    \mathcal{V} = \left(\frac{2r_\mathrm{c}}{D_\mathrm{c}}\right)^3.
\end{align}

For sufficiently short time intervals $t$, particles move in rectilinear fashion, and the typical proper displacement of a relativistic particle within $t$ is $r = ct$, not $r = \sqrt{6Dt}$.
Diffusion can only possibly provide an accurate description of the typical displacement for sufficiently large $t$.
As superluminal motion is impossible,
\begin{align}
    \sqrt{6Dt} < ct,\ \text{or } t > \frac{6D}{c^2} \eqqcolon \tau_\mathrm{d},
\end{align}
where $\tau_\mathrm{d}$ is the diffusion timescale (as in \citet{Globus12008}, but with a factor 6 instead of 4).
Diffusion only has a role to play in the description of particle movement through voids when $\tau_\mathrm{d} < \tau_\mathrm{b}$, the ballistic timescale for particle movement through voids.
We define
\begin{align}
    \tau_\mathrm{b} \coloneqq \frac{R_\mathrm{c}}{(1+z)c},
\end{align}
where $R_\mathrm{c}$ is the comoving void radius.
As $\tau_\mathrm{d} \propto D$, there is a maximum diffusion coefficient, $D_\mathrm{max}$, above which the diffusive description is invalid.
Solving $\tau_\mathrm{d}(D_\mathrm{max}) = \tau_\mathrm{b}$ for $D_\mathrm{max}$, we obtain
\begin{align}
    D_\mathrm{max} = \frac{R_\mathrm{c} c}{6 (1+z)}.
\end{align}
This maximum diffusion coefficient corresponds to a minimum magnetic field strength, $B_\mathrm{min}$.
Approximating $D_\mathrm{max} \approx D_\mathrm{Bohm}(E,B_\mathrm{min})$, we find
\begin{align}
    B_\mathrm{min} &= 2\sqrt{\frac{2}{3}}\frac{(1+z)E}{c |q| R_\mathrm{c}}\\
    &= 8.8 \cdot 10^{-20}\ \mathrm{G} \cdot \frac{E}{1\ \mathrm{GeV}} \cdot \frac{20\ \mathrm{Mpc}}{R_\mathrm{c}} \cdot \frac{1+z}{1} \cdot \frac{e}{|q|}.
\end{align}
We should only apply diffusion theory to the problem of particle movement through voids for void magnetic field strengths $B \gg B_\mathrm{min}$.
For particle energies of $1\ \mathrm{GeV}$, we therefore only consider diffusion for $B \gtrsim 10^{-18}\ \mathrm{G}$.

The diffusing cosmic rays lose energy over time.
In voids, losses by inverse Compton scattering to CMB photons are by far more important than losses by synchrotron radiation, because
\begin{align}
    \frac{P_\mathrm{IC,CMB}}{P_\mathrm{s}} = \frac{B_\mathrm{CMB}^2(z)}{B^2},
\end{align}
and $B_\mathrm{CMB}^2(z) \gg B^2$ in voids.
(Here, $P_\mathrm{IC}$ and $P_\mathrm{s}$ are, respectively, the inverse Compton and synchrotron powers of a single cosmic ray.)
The inverse Compton loss timescale for an electron or positron of total energy $E$ at cosmological redshift $z$ is
\begin{align}
    \tau_\mathrm{IC,CMB}(E,z) \coloneqq& \frac{E}{P_\mathrm{IC,CMB}(E,z)}\\
    =& \frac{6m_e^2c^3\mu_0}{4\beta^2E\sigma_\mathrm{T}B^2_\mathrm{CMB}(0)(1+z)^4}\\
    =& 1.2\ \mathrm{Gyr} \cdot \frac{1}{\beta^2} \cdot \frac{1\ \mathrm{GeV}}{E} \cdot \frac{1}{(1+z)^4},
\end{align}
where $\sigma_\mathrm{T}$ is the Thomson cross-section for electrons and positrons.
For protons, the inverse Compton loss timescale equals the above multiplied by a factor $\left(\frac{m_p}{m_e}\right)^4 \approx 1.1 \cdot 10^{13}$.
Therefore, for non--ultra-high-energy cosmic ray protons, both synchrotron \emph{and} inverse Compton losses are negligible.

Because $\frac{\mathrm{d}D}{\mathrm{d}E} > 0$, diffusion slows down as particles lose energy; in other words, a particle's \emph{highest} diffusion coefficient is its \emph{initial} diffusion coefficient.

Let $X_\mathrm{c}$ be the comoving displacement along the $x$-direction.
We consider $N$ time steps, each of length $\tau \coloneqq \frac{t}{N}$.
Let $X_{\mathrm{c},i}$ be the comoving displacement along the $x$-direction achieved in the $i$-th time step, and let $X_i$ be the corresponding proper displacement.
\begin{align}
    X_\mathrm{c} \coloneqq \sum_{i=1}^N X_{\mathrm{c},i}
\end{align}
Because $\mathbb{E}[X_{\mathrm{c},i}] = 0$, $\mathbb{E}[X_\mathrm{c}] = \sum_{i=1}^N \mathbb{E}[X_{\mathrm{c},i}] = 0$.
Therefore
\begin{align}
\mathbb{E}[X_\mathrm{c}^2] = \mathbb{V}[X_\mathrm{c}] = \sum_{i=1}^N \mathbb{V}[X_{\mathrm{c},i}] = \sum_{i=1}^N \mathbb{E}[X_{\mathrm{c},i}^2].
\end{align}
Because $X_{\mathrm{c},i} = (1+z_i)X_i$, $\mathbb{E}[X_{\mathrm{c},i}^2] = (1+z_i)^2 \mathbb{E}[X_i^2]$.
Therefore
\begin{align}
\mathbb{E}[X_\mathrm{c}^2] = \sum_{i=1}^N (1+z_i)^2 \frac{\mathbb{E}[X_i^2]}{2\tau} \cdot 2\tau.
\end{align}
Following Einstein's definition of the diffusion coefficient, the proper diffusion coefficient in the $x$-direction for the $i$-th time step, $D_{x,i}$, is
\begin{align}
    D_{x,i} \coloneqq \frac{\mathbb{E}[X_i^2]}{2\tau}.
\end{align}
We can then write
\begin{align}
    \mathbb{E}[X_\mathrm{c}^2] = 2\tau \sum_{i=1}^N (1+z_i)^2 D_{x,i}.
\end{align}
Proceeding analogously for the $y$- and $z$-directions, and defining $R_\mathrm{c}^2 \coloneqq X_\mathrm{c}^2 + Y_\mathrm{c}^2 + Z_\mathrm{c}^2$, we find
\begin{align}
    \mathbb{E}[R_\mathrm{c}^2] &= \mathbb{E}[X_\mathrm{c}^2] + \mathbb{E}[Y_\mathrm{c}^2] + \mathbb{E}[Z_\mathrm{c}^2]\\
    &= 2\tau \sum_{i=1}^N (1+z_i)^2 (D_{x,i} + D_{y,i} + D_{z,i}).
\end{align}
In the isotropic case, $D_{x,i} = D_{y,i} = D_{z,i} \eqqcolon D_i$, so that
\begin{align}
    \mathbb{E}[R_\mathrm{c}^2] = 6\tau \sum_{i=1}^N (1+z_i)^2 D_i = 6t \cdot \frac{1}{N} \sum_{i=1}^N (1+z_i)^2 D_i.
\end{align}
Denoting the (time) average of a function $f(t)$ by $\langle f \rangle$, we have
\begin{align}
    r_\mathrm{c} \coloneqq \sqrt{\mathbb{E}[R_\mathrm{c}^2]} = \sqrt{6 \langle (1+z)^2 D \rangle t}.
\end{align}

Let $E$ be an RV denoting particle energy, and let $f_E$ be its PDF.
Let $n_\mathrm{l}$ be the lobe particle number density, and let $R_\mathrm{l}$ be the lobe radius.
The number of particles in an outer shell with thickness $\Delta R$ with energies between $E$ and $E + \mathrm{d}E$ is
\begin{align}
    \mathrm{d}N(E) = 4\pi R_\mathrm{l}^2 \Delta R \cdot n_\mathrm{l} f_E(E)\mathrm{d}E.
\end{align}
These particles escape from the shell over a timescale
\begin{align}
    \tau_\mathrm{e}(E) = \frac{\Delta R^2}{2 D_{\perp,\mathrm{c}}(E)},
\end{align}
where $D_{\perp,\mathrm{c}}$ is the compound cross-field diffusion coefficient \citep{Ensslin11999}.
The number of particles with energies between $E$ and $E + \mathrm{d}E$ escaping from the shell per unit of time thus is
\begin{align}
    \frac{\mathrm{d}N(E)}{\tau_\mathrm{e}(E)} = \frac{8\pi R_\mathrm{l}^2 \cdot D_{\perp,\mathrm{c}}(E) \cdot n_\mathrm{l} f_E(E)\mathrm{d}E}{\Delta R}.
\end{align}
The particulate energy escaping from the shell per unit of time and unit of energy, which we shall call the power density $P_E$, is
\begin{align}
    P_E(E) \coloneqq E \cdot \frac{\mathrm{d}N(E)}{\tau_\mathrm{e}(E) \mathrm{d}E} = \frac{8\pi R_\mathrm{l}^2 \cdot D_{\perp,\mathrm{c}}(E) \cdot n_\mathrm{l} f_E(E)E}{\Delta R}.
\end{align}
Finally, the total power $P$ is
\begin{align}
    P \coloneqq \int_E P_E(E)\ \mathrm{d}E &= \frac{8\pi R_\mathrm{l}^2 n_\mathrm{l}}{\Delta R} \int_E D_{\perp,\mathrm{c}}(E) E f_E(E)\ \mathrm{d}E\\
    &= \frac{8\pi R_\mathrm{l}^2 n_\mathrm{l}}{\Delta R} \mathbb{E}_E[D_{\perp,\mathrm{c}}(E) E].
\end{align}
The compound cross-field diffusion coefficient is \citep{Duffy11995}
\begin{align}
    D_{\perp,\mathrm{c}}(E) \approx D_\perp(E) \left(1 + \frac{\Lambda^2(E)}{\ln{\Lambda(E)}}\right),
\end{align}
where $D_\perp$ is the cross-field diffusion coefficient, given by
\begin{align}
    D_\perp(E) \approx \frac{c}{3}r_L(E) \delta_B(r_L(E)).
\end{align}
Here
\begin{align}
    \delta_B(l) \approx f_\mathrm{i} \cdot \left(\frac{l}{l_\mathrm{i}}\right)^{\frac{2}{3}},
\end{align}
where $l_\mathrm{i}$ is the turbulence injection scale and $f_\mathrm{i}$ is the total turbulence energy density up to this scale, relative to the energy density of the thermal medium surrounding the lobe \citep{Ensslin11999}.
Additionally,
\begin{align}
    \Lambda(E) = \frac{1}{\sqrt{2}} \frac{\delta_B(\lambda_\mathrm{c,l})}{\delta_B(r_\mathrm{L}(E))} = \frac{1}{\sqrt{2}} \left(\frac{\lambda_{\mathrm{c,l}}}{r_\mathrm{L}(E)}\right)^{\frac{2}{3}},
\end{align}
where $\lambda_\mathrm{c,l}$ is the lobe's magnetic field correlation length.
We assume that $E$ has a Pareto distribution \citep[e.g.][]{Werner12016}, so that its PDF, $f_E$, (for $p \neq -1$) is given by
\begin{align}
    f_E(E) = \begin{cases}
        \frac{p+1}{E_\mathrm{max}^{p+1} - E_\mathrm{min}^{p+1}} E^p & \text{if } E_\mathrm{min} < E < E_\mathrm{max};\\
        0 & \text{otherwise};
    \end{cases}
\end{align}
and $E_\mathrm{min} \coloneqq \gamma_\mathrm{min} m c^2$ and $E_\mathrm{max} \coloneqq \gamma_\mathrm{max} m c^2$.
We calculated the total power assuming $R_\mathrm{l} = 100\ \mathrm{kpc}$, $\Delta R = \lambda_\mathrm{c,l} = 10\ \mathrm{kpc}$, $l_\mathrm{i} = 10\ \mathrm{kpc}$, $f_\mathrm{i} = 10^{-2}$, $B = B_\mathrm{l} = 10^{-7}\ \mathrm{G}$, $|q| = e$, $m = m_e$, $p = -2.4$, $\gamma_\mathrm{min} = 10$, $\gamma_\mathrm{max} = 10^5$, and $n_\mathrm{l} = 10^{-10}\ \mathrm{cm}^{-3}$.
We find $P = 10^{30}\ \mathrm{W}$.

To estimate the final void magnetic field strength $B_\mathrm{v}$, we followed an argument akin to that in \citet{Beck12013}.
If the lobe would expand to fill the entire void, then magnetic flux conservation yields
\begin{align}
    B_\mathrm{v} = B_\mathrm{l} \left(\frac{R_\mathrm{l}}{R_\mathrm{v}}\right)^2.
\end{align}
By squaring and dividing both sides of this equation by $2\mu_0$, one recasts it in terms of magnetic energy densities and obtains
\begin{align}
    u_{B_\mathrm{v}} = u_{B_\mathrm{l}} \left(\frac{R_\mathrm{l}}{R_\mathrm{v}}\right)^4.
\end{align}
However, only a fraction of the lobe's magnetic energy can escape, and it is only this fraction that we should consider in our calculation.
If we assume that the magnetic energy that is carried out of the lobe is comparable to the energy of the escaped particles, which equals $Pt$, then
\begin{align}
    u_{B_\mathrm{v}} = \frac{Pt}{E_\mathrm{l}} u_{B_\mathrm{l}} \left(\frac{R_\mathrm{l}}{R_\mathrm{v}}\right)^4,
\end{align}
where $E_\mathrm{l}$ is the total magnetic energy of the lobe.
Recasting this equation back to magnetic field strengths, we obtain
\begin{align}
    B_\mathrm{v} = \sqrt{\frac{Pt}{E_\mathrm{l}}}\left(\frac{R_\mathrm{l}}{R_\mathrm{v}}\right)^2 B_\mathrm{l}.
\end{align}
The energy ratio in \citet{Beck12013}'s analogous Eq.~4 should likewise appear under a square root.
This is a matter of typo\-graphy only: the authors did take the square root to obtain their results (private communication with M. Hanasz).
For $t = 10^0\ \mathrm{Gyr}$, $E_\mathrm{l} = 10^{55}\ \mathrm{J}$, $R_\mathrm{v} = 10^1\ \mathrm{Mpc}$, and $P$, $R_\mathrm{l}$, and $B_\mathrm{l}$ as before, we obtained $B_\mathrm{v} = 6 \cdot 10^{-16}\ \mathrm{G}$.

\end{appendices}

\end{document}